\documentclass[twoside,twocolumn,english,prc,amsmath,a4paper,myheadings]{revtex4}
\usepackage[T1]{fontenc}
\usepackage{geometry}
\usepackage{amsmath}
\usepackage{graphicx}
\usepackage{amssymb}

\makeatletter
\def\@oddhead{\rightmark \hfill  Jets, Bulk Matter, and their Interaction  \hfill \thepage}
\def\@evenhead{\thepage \hfill K. Werner et al.\hfill}
\def\fnum@table{\tablename~{\bf\thetable}}
\def\fnum@figure{\figurename~{\bf\thefigure}}
\def\tablename{\footnotesize{\bf Table}}
\def\figurename{\footnotesize{\bf Figure}}

\usepackage{dcolumn}

\def\citet{\cite}

\makeatother

\usepackage{babel}

\begin{document}

\title{Jets, Bulk Matter, and their Interaction in Heavy Ion Collisions
at Several TeV}

\author{{\normalsize K.$\,$Werner$^{(a)}$, Iu.$\,$Karpenko$^{(b)}$, M.
Bleicher$^{(c)}$, T.$\,$Pierog$^{(d)}$, S. Porteboeuf-Houssais$^{(e)}$ }}

\address{$^{(a)}$ SUBATECH, University of Nantes -- IN2P3/CNRS-- EMN, Nantes,
France}

\address{$^{(b)}$ Bogolyubov Institute for Theoretical Physics, Kiev 143,
03680, Ukraine}

\address{$^{(c)}$ FIAS, Johann Wolfgang Goethe Universitaet, Frankfurt am
Main, Germany}

\address{$^{(d)}$Karlsruhe Inst. of Technology, KIT, Campus North, Inst.
f. Kernphysik, Germany}

\address{$^{(e)}$LPC, Université Blaise Pascal, CNRS-IN2P3, Clermont-Ferrand,
France}

\begin{abstract}
We discuss a theoretical scheme that accounts for bulk matter, jets,
and the interaction between the two. The aim is a complete description
of particle production at all transverse momentum ($p_{t}$) scales.
In this picture, the hard initial scatterings result in mainly longitudinal
flux tubes, with transversely moving pieces carrying the $p_{t}$
of the partons from hard scatterings. These flux tubes constitute
eventually both bulk matter (which thermalizes and flows) and jets.
We introduce a criterion based on parton energy loss to decide whether
a given string segment contributes to the bulk or leaves the matter
to end up as a jet of hadrons. Essentially low $p_{t}$ segments from
inside the volume will constitute the bulk, high $p_{t}$ segments
(or segments very close to the surface) contribute to the jets. The
latter ones appear after the usual flux tube breaking via q-qbar production
(Schwinger mechanism). Interesting is the transition region: Intermediate
$p_{t}$ segments produced inside the matter close to the surface
but having enough energy to escape, are supposed to pick up q-qbar
pairs from the thermal matter rather than creating them via the Schwinger
mechanism. This represents a communication between jets and the flowing
bulk matter (fluid-jet interaction). Also very important is the interaction
between jet hadrons and the soft hadrons from the fluid freeze-out.
We employ the new picture to investigate Pb-Pb collisions at 2.76
TeV. We discuss the centrality and $p_{t}$ dependence of particle
production  and long range dihadron correlations at small and large
$p_{t}$.
\end{abstract}
\maketitle

\section{Introduction}

Traditionally the physics of ultrarelativistc heavy ion collisions
is discussed in terms of different categories like collective dynamics,
parton-jet physics, and fluctuation-correlation studies, although
these different topics are highly correlated. In this article, a complete
dynamical picture of particle production at all $p_{t}$ scales will
be presented, which accounts for the production and evolution of bulk
matter and jets, and the very important interaction between the two
components (which is not only the well known parton energy loss).
The consequences of these interactions can be nicely seen in long
range dihadron correlations.

The physical picture of our approach is the following: Initial hard
scatterings result in mainly longitudinal flux tubes, with transversely
moving pieces carrying the $p_{t}$ of the partons from hard scatterings.
These flux tubes constitute eventually both bulk matter (which thermalizes,
flows, and finally hadronizes) and jets, according to some criteria
based on partonic energy loss. We will consider a sharp fluid freeze-out
hypersurface, defined by a constant temperature. Freeze-out here simply
means the end of the fluid phase, but the hadrons still interact.
High energy flux tube segments will leave the fluid, providing jet
hadrons via the usual Schwinger mechanism of flux-tube breaking caused
by quark-antiquark production. 

But the jets may also be produced at the freeze-out surface. Here
we assume that the quark-antiquark needed for the flux tube breaking
is provided by the fluid, with properties (momentum, flavor) determined
by the fluid rather than the Schwinger mechanism. Considering transverse
fluid velocities up to 0.7c, and thermal parton momentum distributions,
one may get a {}``push'' of a couple of GeV to be added to the transverse
momentum of the string segment. This will be a crucial effect for
intermediate $p_{t}$ jet hadrons. 

There is another important issue. Even for hadrons with transverse
momenta of 10-20 GeV, there is a large probability of a jet hadron
formation before it enters the dense hadronic medium. This means a
significant probability of scatterings between jet hadrons and soft
hadrons (from freeze-out), having essentially two consequences: an
increase of low $p_{t}$ particle production, and a reduction of yields
at high $p_{t}.$ In addition there are of course the well known hadronic
interactions between the soft hadrons.

We have discussed different processes which all affect $p_{t}$ spectra.
It is, however, possible to disentangle different contributions, by
looking at dihadron correlations. These are extremely useful tools
heavily used by experimental groups at the RHIC and the LHC \citet{ridgeSTAR,ridgePHEN,ridgePHOB,ridgeCMS,ridgeALI}.
Recently, the CMS and ALICE collaborations published results on such
correlations in Pb-Pb collisions at 2.76 GeV and different centralities,
over a more or less broad range in relative pseudorapidity ($\Delta\eta$)
and full coverage of the relative azimuthal angle ($\Delta\phi$)
\citet{ridgeCMS,ridgeALI}. Different combinations of transverse momenta
$p_{t}^{\mathrm{assoc}}$ and $p_{t}^{\mathrm{trigg}}$ of associated
and trigger particles in the range between 0.25 GeV/c and 15 GeV/c
are investigated. Considering long range correlations ($|\Delta\eta|>A,\; A\ge0.8$),
the coefficients $V_{n\Delta}$ of the harmonic decomposition factorize
as $V_{n\Delta}=v(p_{t}^{\mathrm{assoc}})\, v(p_{t}^{\mathrm{trigg}}$)
-- not only for small transverse momenta but also for example for
large $p_{t}^{\mathrm{trigg}}$ and small $p_{t}^{\mathrm{assoc}}$.
For small momenta the situation seems to be clear: the correlation
is flow dominated. But factorization does not necessarily mean that
both hadrons carry the flow from the fluid! This can in particular
not be the answer for observed correlations for large $p_{t}^{\mathrm{trigg}}$
-- here we have to deal with an interaction between the flowing bulk
and jets, which makes the observed correlations very interesting in
particular as a test of our ideas concerning bulk-jet separation and
interaction.

Another challenge : The ATLAS collaboration showed recently results
\citet{v2ATLAS} on elliptical flow of charged particles with respect
to an event plane in the opposite $\eta$ hemisphere (also a kind
of long range correlation). The $v_{2}$ values are quite large up
to values of $p_{t}=20$GeV/c, for eight different centrality ranges.
Can we understand this in a quantitative fashion?

The heavy ion results shown in this paper are based on 2000000 events
simulated with EPOS2.17v3. A central (0-5\%) Pb-Pb event takes on
the average around 2 HS06 hours CPU time, using machines with an average
scaling factor of 8.7 \citet{hs06}. Always six events share the same
parton configuration and hydrodynamic evolution, with only particle
production and hadronic rescattering being redone (to gain statistics).
This is taken care of when considering mixed events in correlation
studies.

\section{The basis: Flux tubes from a multiple scattering approach}

\begin{figure}[tb]
\begin{centering}
\includegraphics[scale=0.21]{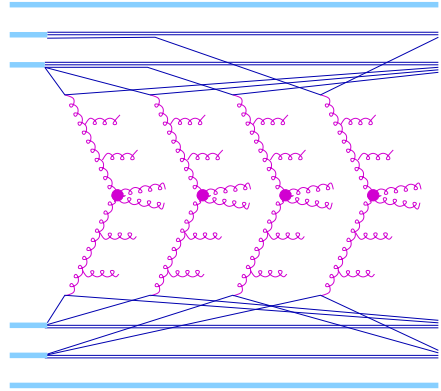}
\par\end{centering}

\caption{(Color online) Multiple scattering in nucleus-nucleus collisions.\label{ldmultaa.eps}}

\end{figure}

\begin{figure}[tb]
\begin{centering}
\includegraphics[scale=0.35]{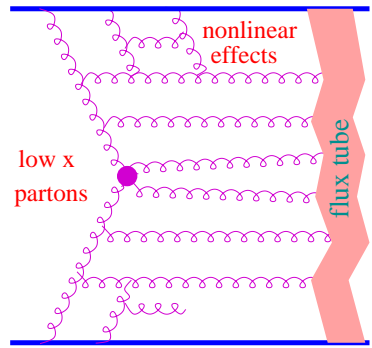}
\par\end{centering}

\caption{(Color online) An elementary parton ladder, whose final state is identified
with a color flux tube (kinky string).\label{ldfluxaa7a.eps}}

\end{figure}

The starting point is a multiple scattering approach corresponding
to a marriage of Gribov-Regge theory and perturbative QCD (pQCD),
see Fig. \ref{ldmultaa.eps}. %
An elementary scattering corresponds to a parton ladder, containing
a hard scattering calculable based on pQCD, including initial and
final state radiation (for details see \citet{epos2}). These ladders
are identified with flux tubes (see Fig. \ref{ldfluxaa7a.eps}), %
which are mainly longitudinal objects, with transversely moving parts,
carrying the transverse momenta of the hard scatterings. These objects
are also referred to as kinky strings. One should note that here multiple
scattering does not mean just a rescattering of hard partons, it rather
means a multiple exchange of complete parton ladders, leading to many
flux tubes. In this case, the energy sharing between the different
scatterings will be very important, to be discussed later. 

The consistent quantum mechanical treatment of the multiple scattering
is quite involved, it is based on cutting rule techniques to obtain
partial cross sections, which are then simulated with the help of
Markov chain techniques \citet{hajo}.

As said before, the final state partonic system corresponding to elementary
parton ladders are identified with flux tubes. The relativistic string
picture \citet{string1,string2,string3} is very attractive, because
its dynamics is essentially derived from general principles as covariance
and gauge invariance. The simplest possible string is a surface $X(\alpha,\beta)$
in 3+1 dimensional space-time, with piecewise constant initial velocities
$\partial X/\partial\beta$. These velocities are identified with
parton velocities,%
\begin{figure}[tb]
\begin{raggedright}
{\large \hspace*{1.2cm}(a)}
\par\end{raggedright}{\large \par}

\vspace*{-0.5cm}

\begin{centering}
\includegraphics[scale=0.15]{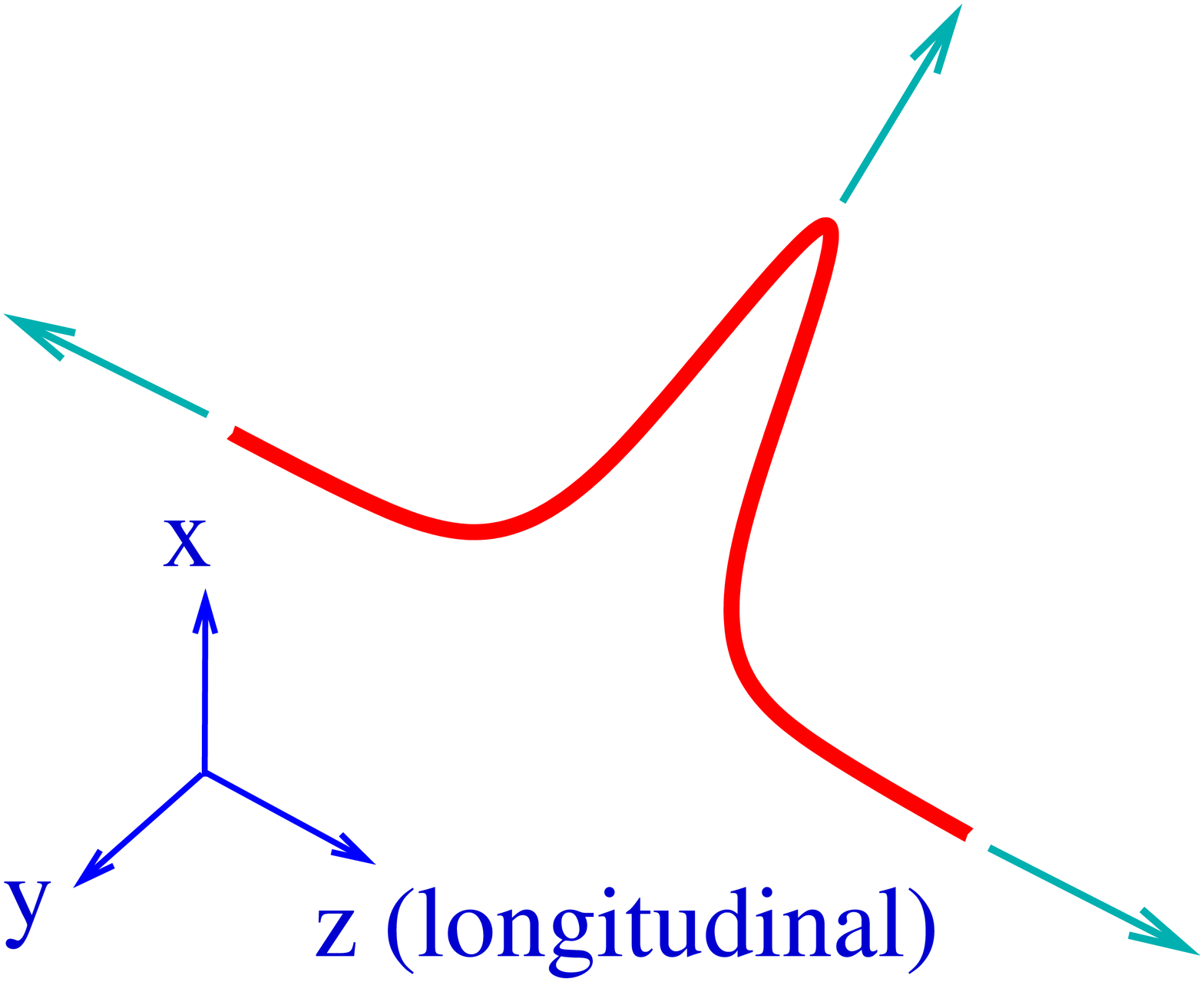}
\par\end{centering}

\begin{raggedright}
{\large \hspace*{1.2cm}(b)}
\par\end{raggedright}{\large \par}

\vspace*{-0.5cm}

\begin{centering}
\includegraphics[scale=0.15]{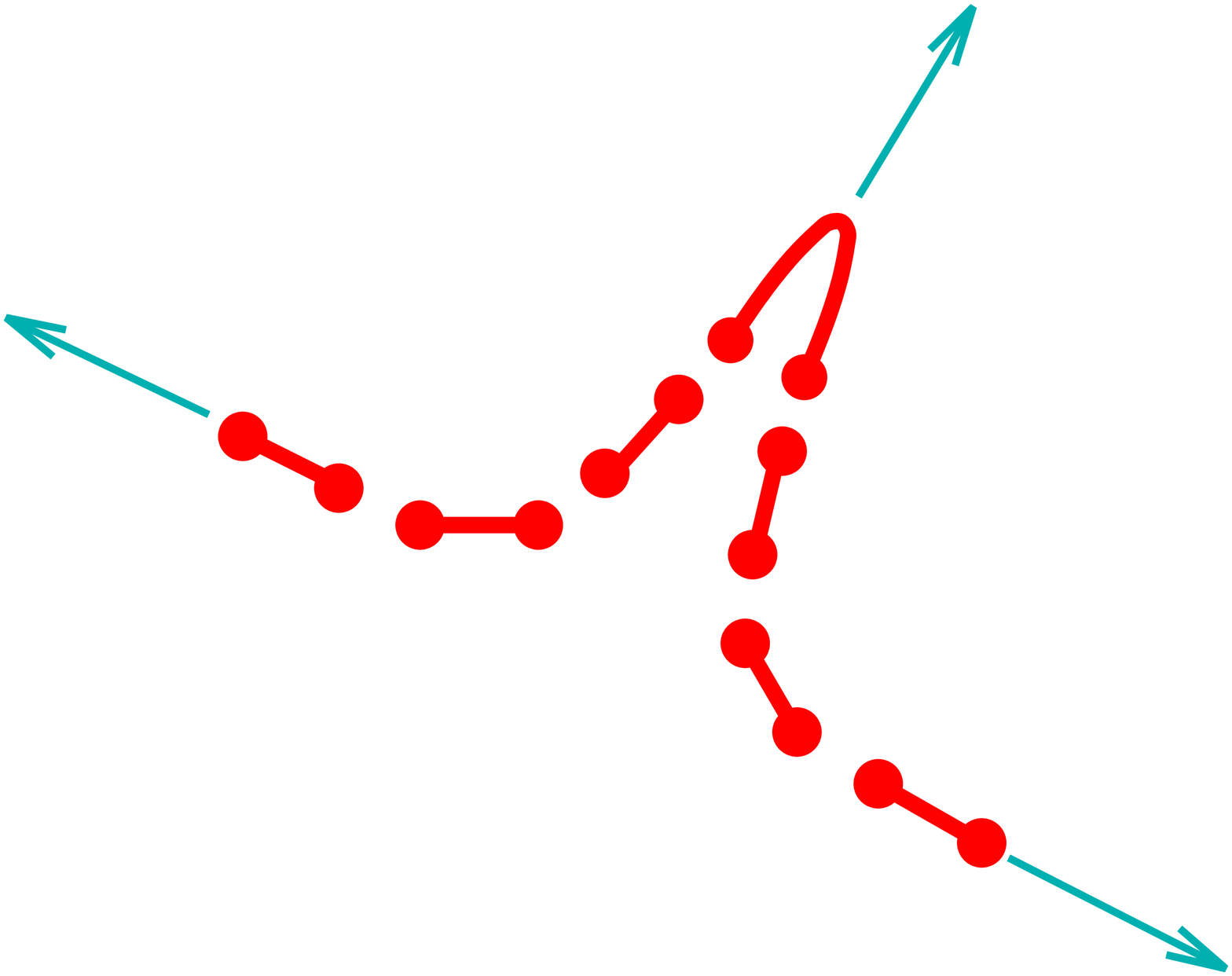}
\par\end{centering}

\caption{(Color online) (a) Flux tube with transversely moving part (kinky
string) in space, at given proper time. \protect \\
(b) Flux tube breaking via $q-\bar{q}$ production, which screens
the color field (Schwinger mechanism).\label{stgfrag1.eps}}

\end{figure}
 which provides a one to one mapping from partons to strings. For
details see \citet{hajo,epos2}. The high transverse momentum ($p_{t})$
partons will show up as transversely moving string pieces, see Fig.
\ref{stgfrag1.eps}(a). Despite the fact that in the TeV energy range
most processes are hard, and despite the theoretical importance of
very high $p_{t}$ partons, it should not be forgotten that the latter
processes are rare, most kinks carry only few GeV of transverse momentum,
and the energy is nevertheless essentially longitudinal. In case of
elementary reactions, the strings will break (see Fig. \ref{stgfrag1.eps}(b)
via the production of quark-antiquark pairs according to the so-called
area law \citet{artru2,mor87,hajo,epos2}. The string segments are
identified with final hadrons and resonances. 

\begin{figure}[b]
\begin{centering}
\includegraphics[angle=270,scale=0.26]{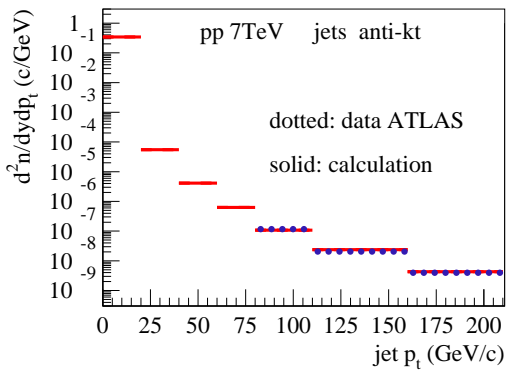}
\par\end{centering}

\caption{(Color online) Inclusive $p_{t}$ distribution of jets. We show the
calculation (full lines) compared to ATLAS data \citet{atlasjet}
(dotted lines).\label{z-seljet-1.ps}}

\end{figure}
\begin{figure}[tb]
\begin{centering}
\includegraphics[angle=270,scale=0.26]{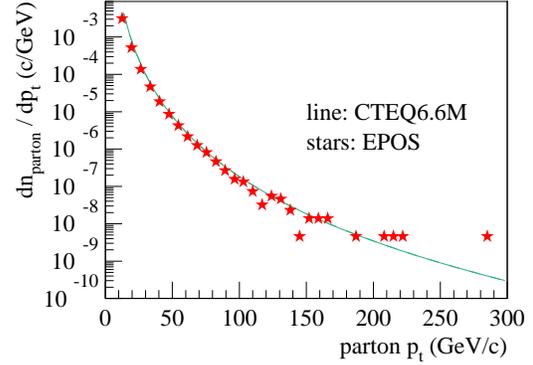}
\par\end{centering}

\caption{(Color online) Transverse momentum distribution of partons: We compare
our results (red stars) with a parton model calculation (dotted line).\label{z-seljet-4.ps}}

\end{figure}
This picture has been very successful to describe particle production
in electron-positron annihilation or in proton-proton scattering at
very high energies. In the latter case, not only low $p_{t}$ particles
are described correctly, for example for pp scattering at 7 TeV \citet{kw1,kw2},
also jet production is covered. As discussed earlier, the high transverse
momenta of the hard partons show up as kinks, transversely moving
string regions. After string breaking, the string pieces from these
transversely moving areas represent the jets of particles associated
with the hard partons. To demonstrate that this picture also works
quantitatively, we compute the inclusive $p_{t}$ distribution of
jets, reconstructed with the anti-kt algorithm \citet{antikt} and
compare with data \citet{atlasjet}, see Fig. \ref{z-seljet-1.ps}.
In Fig. \ref{z-seljet-4.ps}, we compare our $p_{t}$ distribution
of partons with a parton model calculation based on CTEQ6 parton distribution
functions \citet{cteq6}, in both cases leading order with a K-factor
of 2.

\section{Jet -- bulk separation}

In heavy ion collisions and also in high multiplicity events in proton-proton
scattering at very high energies, the density of strings will be so
high that the strings cannot decay independently as described above.
Here we have to modify the procedure as discussed in the following.
The starting point are still the flux tubes (kinky strings) originating
from elementary collisions. These flux tubes will constitute both,
bulk matter which thermalizes and expands collectively, and jets.
The criterion which decides whether a string piece ends up as bulk
or jet, will be based on energy loss. In the following we consider
a flux tube in matter, where {}``matter'' first means the presence
of a high density of other flux tubes, which then thermalize. A more
quantitative discussion will follow.

\begin{figure}[b]
\begin{raggedright}
{\large \hspace*{1.2cm}(a)}
\par\end{raggedright}{\large \par}

\vspace*{-0.5cm}

\begin{centering}
\includegraphics[scale=0.15]{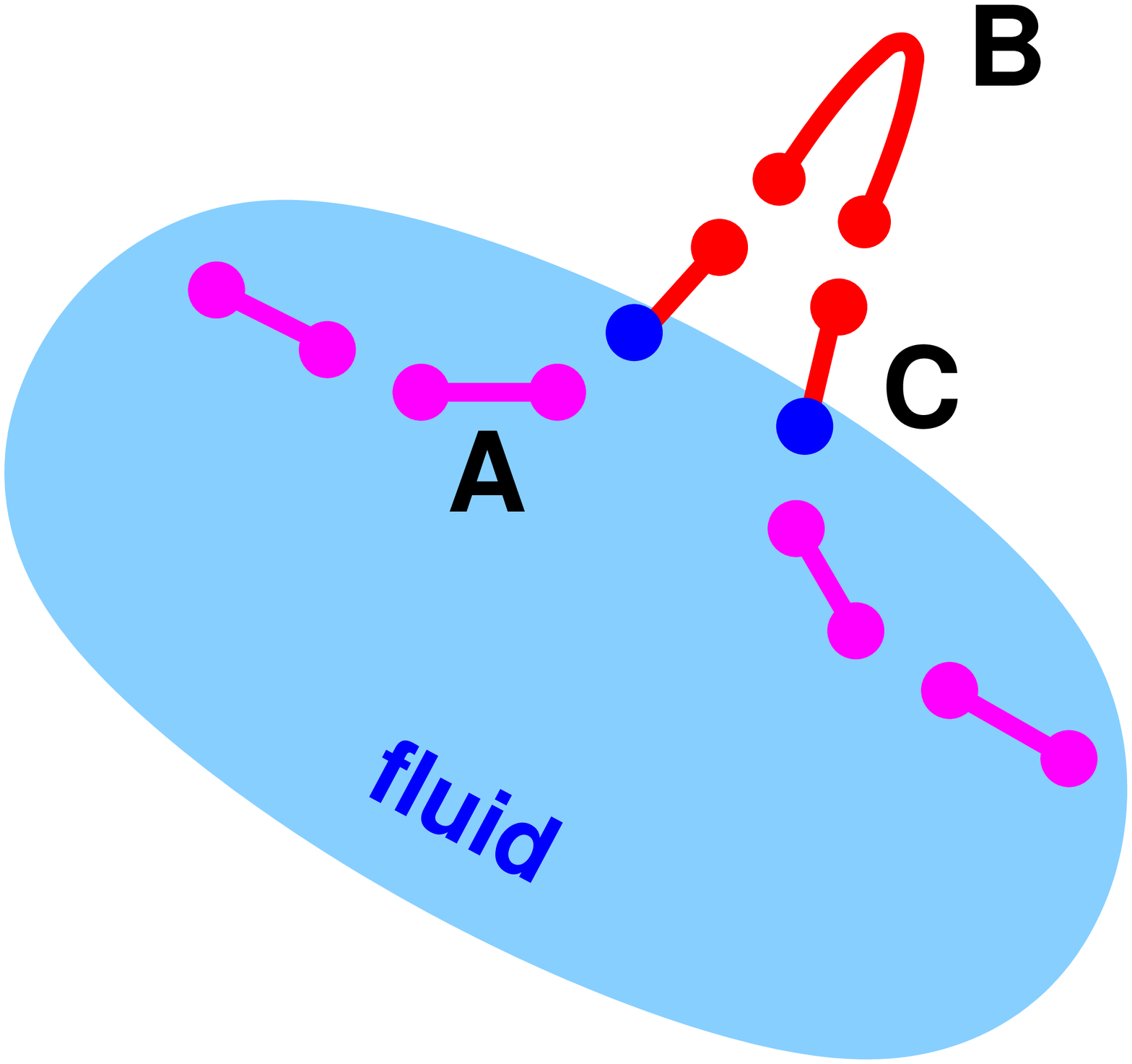}
\par\end{centering}

\begin{raggedright}
{\large \hspace*{1.2cm}(b)}
\par\end{raggedright}{\large \par}

\vspace*{-0.5cm}

\begin{centering}
\includegraphics[scale=0.15]{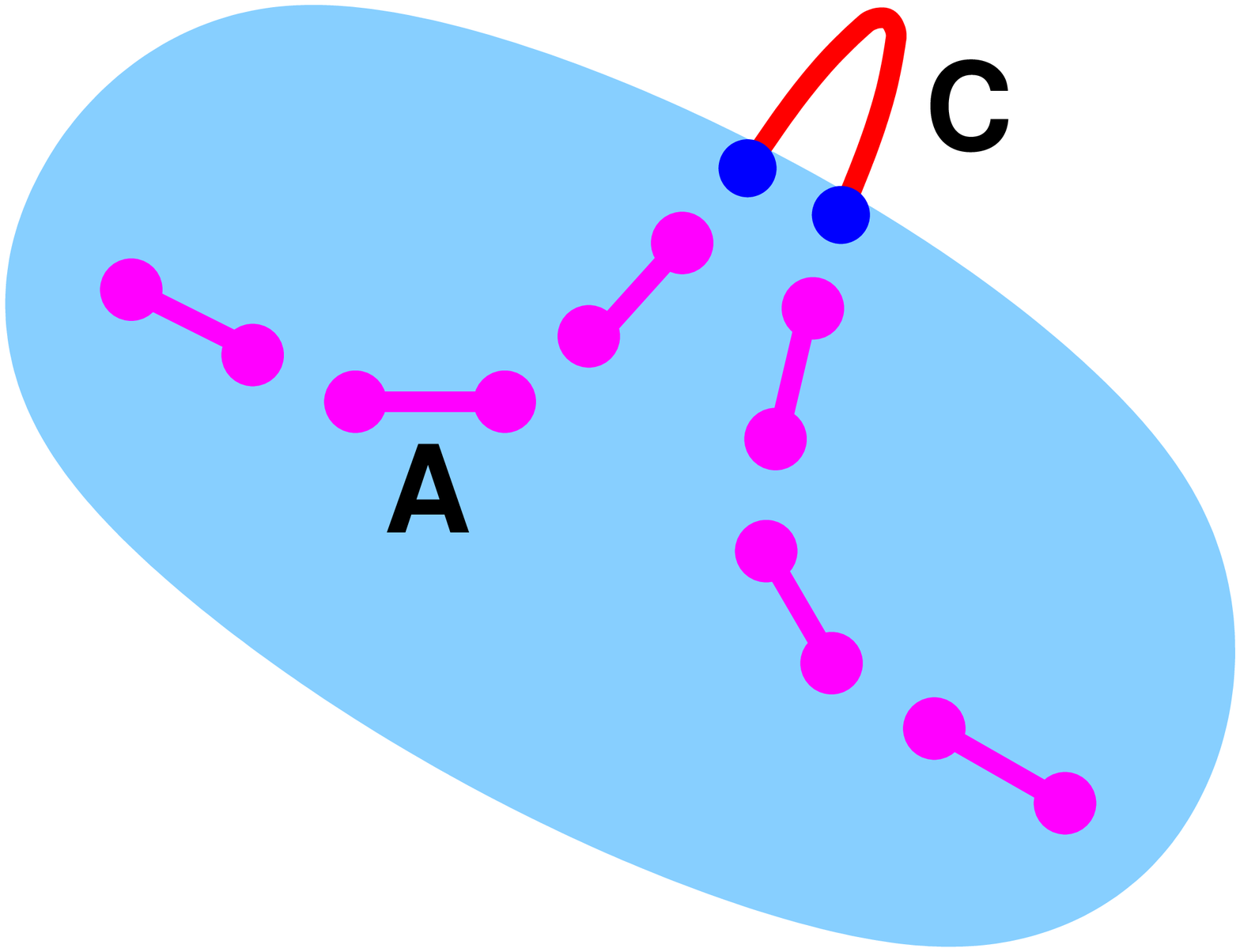}
\par\end{centering}

\caption{(Color online) Flux tube in matter (from other flux tubes, blue colored
area). One distinguishes three types of behavior for string segments,
noted as A, B, C (see text). \label{stgfrag3.eps} The highest $p_{t}$
string segment may be (a) of type B or (b) of type C.}

\end{figure}

Three possibilities should occur, referred to as A, B, C, see Fig.
\ref{stgfrag3.eps} :

\begin{description}
\item [{A}] String segments far from the surface and/or being slow will
simply constitute matter, they loose their character as individual
strings. This matter will evolve hydrodynamically and finally hadronize
({}``soft hadrons''). 
\item [{B}] Some string pieces (like those close to transversely moving
kinks) will be formed outside the matter, they will escape and constitute
jets ({}``jet hadrons''). 
\item [{C}] There are finally also string pieces produced inside matter
or at the surface, but having enough energy to escape and show up
as jets ({}``jet hadrons''). They are affected by the flowing matter
({}``fluid-jet interaction''). 
\end{description}
Let us discuss how the above ideas are realized. In principle the
formation and expansion of matter and the interaction of partons with
matter is a dynamical process. However, the initial distribution of
energy density and the knowledge of the initial momenta of partons
(or string segments) allows already an estimate about the fate of
the string segments. By {}``initial time'' we mean some early proper
time $\tau_{0}$ which is a parameter of the model. Strictly speaking,
energy loss concerns partons, modifying eventually the kink momenta
in our picture, and the momenta of the string segments after breaking
will be reduced. We will therefore base our discussion on energy loss
on string segments. 

We estimate the energy loss $\Delta E$ of string segments along their
trajectory to be\begin{equation}
\Delta E=k_{\mathrm{Eloss}}\, E_{0}\int(\rho V_{0})^{3/8}\max(1,\sqrt{E/E_{0}}\,)\, dL/L_{0},\label{eq:eloss}\end{equation}
inspired by \citet{peigne}, where $\rho$ is the density of string
segments at initial proper time $\tau_{0}$, $V_{0}$ is an elementary
volume cell size (technical parameter, taken to be 0.147~fm$^{3}$),
$L_{0}$ is a (technical) length scale (taken to be 1 fm), $E$ the
energy of the segment in the {}``Bjorken frame'' moving with a rapidity
$y$ equal to the space-time rapidity $\eta_{s}$, $dL$ is a length
element, and $k_{\mathrm{Eloss}}$ and $E_{0}$ are parameters. We
introduce an energy cutoff $E_{0}$ to have sufficient energy loss
for slowly moving segments.

A string segment will contribute to the bulk (type A segment), when
its energy loss is bigger than its energy, i.e.\begin{equation}
\Delta E\ge E.\end{equation}
All the other segments are allowed to leave the bulk (type B or C
segments). Only the bulk segments are used to determine the initial
conditions for hydrodynamics, following the same procedure as explained
in \citet{epos2} (with some new elements, as discussed in the next
section). Starting from this initial condition, the bulk matter will
evolve according to the equations of ideal hydrodynamics till {}``hadronization'',
which occurs at some {}``hadronization temperature'' $T_{H}$ \citet{epos2}.
Hadronization means that we change from matter description to particle
description, but hadrons still interact among each other, realized
via a hadronic cascade procedure \citet{urqmd}, already discussed
in \citet{epos2}.

After having performed the hydrodynamic expansion, we have to come
back to the string segments which escape the bulk because their energy
is bigger than the energy loss. We employ a formation time: the string
segments are formed at times $t$ distributed as $\exp(-t/\gamma\tau_{\mathrm{form}})$,
with some parameter $\tau_{\mathrm{form}}$ which is taken to be 1fm/c.
If the formation time is such that the segment is produced outside
the {}``hadronization surface'' defined by $T_{H}$, the segment
will escape as it is (type B segment). 

\begin{figure}[tb]
\begin{centering}
\includegraphics[scale=0.15]{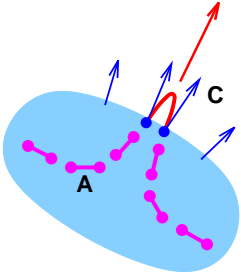}
\par\end{centering}

\caption{(Color online) A type C segment picks up quark and antiquark from
the fluid, carrying momenta and flavor according to the fluid properties
({}``fluid-jet interaction'').  \label{stgfrag8.eps}}

\end{figure}

Most interesting are the segments which are formed inside but still
escape, because they have $E>\Delta E$. These are type C segments.
They escape, but their properties change. Actually such a segment
leaves {}``matter'' at the hadronization surface at a particular
space-time point $x$, which is characterized by some collective flow
velocity $\vec{v}(x)$. We assume that the string breaking in this
case is modified such that the quark and antiquark (or diquark) necessary
for the string breaking are taken from the flowing fluid rather than
being produced via the Schwinger mechanism. So the new string segment
is composed of a quark and antiquark (diquark) carrying the flow velocity,
and the string piece in between, which has not been changed. This
string piece may or may not carry large momentum, depending on whether
it is close to a kink or not, the former possibility shown in Fig.
\ref{stgfrag8.eps}.%

In any case, due to the fluid-jet interaction, the properties of this
segment change drastically compared to the normal fragmentation:

\begin{itemize}
\item The quark and antiquark (or diquark ) from the fluid provide a push
in the direction of the moving fluid.
\item The quark (antiquark) flavors are determined from Bose-Einstein statistics,
with more strangeness production compared to the Schwinger mechanism.
\item The probability $p_{\mathrm{diq}}$ to have a diquark rather than
an antiquark will be bigger compared to a highly suppressed diquark-antidiquark
breakup in the Schwinger picture ($p_{\mathrm{diq}}$ is a parameter).
\end{itemize}
Our procedure has 4 parameters: $k_{\mathrm{Eloss}}$ (=0.042), $E_{0}$
(=6~GeV), $\tau_{\mathrm{form}}$(=1~fm/c), $p_{\mathrm{diq}}$(=0.22).
It allows to cover in a single scheme the production of jets, of bulk,
and the interaction between the two.

\section{Formation times}

\begin{figure}[tb]
\begin{centering}
\includegraphics[angle=270,scale=0.3]{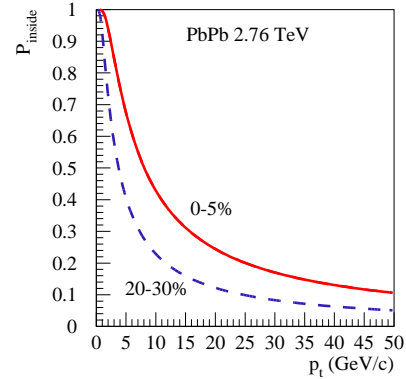}
\par\end{centering}

\caption{(Color online) The estimate $P_{\mathrm{inside}}$ to form (pre)hadrons
inside the fluid, as a function of $p_{t}$, for Pb-Pb collisions
at 2.76 TeV. We show the curves for the 0-5\% and the 20-30\% most
central events.\label{fig:estim2}}

\end{figure}

A crucial ingredient to the mechanism of fluid-jet interaction is
the formation time of jet hadrons (the hadrons which leave the fluid).
The probability distribution of the formation times $t$ of jet hadrons
with gamma factors $\gamma$ is given as \begin{equation}
\mathrm{prob}\sim\exp\left(-\frac{t}{\gamma\tau_{\mathrm{form}}}\right),\label{eq:prob}\end{equation}
(we use $\tau_{\mathrm{form}}=1\,$fm/c). The probability of having
a formation point inside the fluid is obtained as an integral over
eq. (\ref{eq:prob}),\begin{equation}
1-\exp\left(-\frac{t_{\max}}{\gamma\tau_{\mathrm{form}}}\right),\label{eq:iprob}\end{equation}
with $t_{\max}$ being the time corresponding to a formation at the
fluid surface. Rather than making a simulation, we are going to present
a very simple formula providing a rough estimate of the $p_{t}$ dependence
of this probability. For a collision of two Pb nuclei in some centrality
interval, characterized by the mean impact parameter $b$, we use
$ct_{\mathrm{max}}=r_{\mathrm{Pb}}-b/2$, where $r_{\mathrm{Pb}}$
is the radius parameter used in the Wood-Saxon distribution of nucleons.
Considering transversely moving hadrons of mass $m$, we have $\gamma\approx p_{t}/mc$.
The estimate $P_{\mathrm{inside}}$ of the probability to form (pre)hadrons
inside the fluid is\begin{equation}
P_{\mathrm{inside}}=1-\exp\left(-\frac{(r_{\mathrm{Pb}}-b/2)\, m}{p_{t}\,\tau_{\mathrm{form}}}\right).\label{eq:iproba}\end{equation}
In Fig. \ref{fig:estim2}, we %
show the result for the 0-5\% and the 20-30\% most central events
in Pb-Pb collisions at 2.76 TeV, using $c\tau_{\mathrm{form}}=1\,\mathrm{fm}$,
$mc^{2}=1\,\mathrm{GeV}$, $r_{\mathrm{Pb}}=6.5\,\mathrm{fm}$, and
for the average impact parameters \textbf{$b=1.8\,\mathrm{fm}$} (0-5\%)
and \textbf{$b=7.8\,\mathrm{fm}$} (20-30\%). 

By construction, the probability $P_{\mathrm{fluid-jet}}$ of having
a fluid-jet interaction is equal to the probability of forming (pre)hadrons
inside the fluid, so its estimate is given by $P_{\mathrm{inside}}$.
From Fig. \ref{fig:estim2}, we see that the probability is quite
large for intermediate values of $p_{t}$, but even large values (50
GeV/c) are significantly affected. Whether the effect of the interaction
can be seen in some observable is a different question and will be
discussed later.

Several authors have already discussed about {}``in-medium hadronization'',
see for example Ref. \citet{bellwied}, where one also finds an overview
about earlier models on this subject.

\section{Hydrodynamics }

The bulk matter extracted as described above provides the initial
condition for a hydrodynamic evolution. As explained in \citet{epos2},
we compute the energy momentum tensor and the flavor flow vector at
some position $x$ (at $\tau=\tau_{0}$) from the four-momenta of
the bulk string segments. The time $\tau=\tau_{0}$ is as well taken
to be the initial time for the hydrodynamic evolution. This seems
to be a drastic simplification, the justification being as follows:
we imagine to have a purely longitudinal scenario (descibed by flux
tubes) till some proper time $\tau_{\mathrm{flux}}<\tau_{0}$. During
this stage there is practically no transverse expansion, and the energy
per unit of space-time rapidity does not change. This property should
not change drastically beyond $\tau_{\mathrm{flux}}$, so we assume
it will continue to hold during thermalization between $\tau_{\mathrm{flux}}$
and $\tau_{0}$. So although we cannot say anything about the precise
mechanism which leads to thermalization, and therefore we cannot compute
the real $T^{\mu\nu}$, we expect at least the elements $T^{00}$
and $T^{0i}$ to stay close to the flux tube values, and we can use
the flux tube results to compute the energy density. Only $T^{ij}$
will change considerably, but this does not affect our calculation
much.

We employ three-dimensional ideal hydrodynamics as described in \citet{epos2},
with some modification to be discussed in the following. As in \citet{epos2},
we construct the equation of state as \begin{equation}
p=p_{Q}+\lambda\,(p_{H}-p_{Q}),\end{equation}
where $p_{H}$ is the pressure of a resonance gas, and $p_{Q}$ the
pressure of an ideal quark gluon plasma, including bag pressure. We
use an updated $\lambda$: \begin{equation}
\lambda=\exp\left(-z-3z^{2}\right)\Theta(T-T_{c})+\Theta(T_{c}-T),\end{equation}
with\begin{equation}
z=x/(1+x/0.77),\quad x=(T-T_{c})/\delta,\end{equation}
using $\delta=0.24\exp(-\mu_{b}^{2}/0.4^{2})$.%
\begin{figure}[tb]
\begin{centering}
\includegraphics[angle=270,scale=0.26]{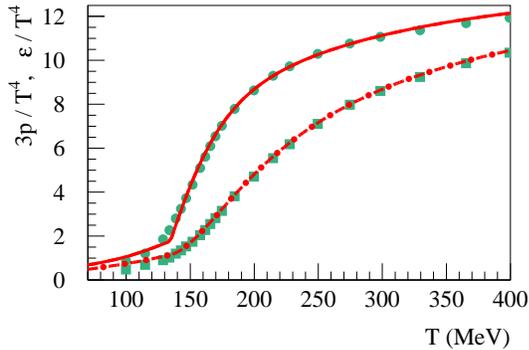}
\par\end{centering}

\caption{(Color online) Energy density and pressure versus temperature, for
our equation-of-state (lines) compared to lattice data \citet{lattice}
(points). \label{cap:eos1}}

\end{figure}
The new $\lambda$ provides an equation of state in agreement with
recent lattice data \citet{lattice}, see Fig. \ref{cap:eos1}. 

Apart of the new equation of state, we use the same procedure to obtain
energy density and pressure from the string segments, as described
in \citet{epos2}. However, 

\begin{itemize}
\item doing the calculation for Pb-Pb collisions at 2.76 TeV, we get too
much elliptical flow (20-30\%), a hint that one should include viscosity. 
\end{itemize}
Taking the usual small radii of the elementary flux-tubes, we get
extremely strongly fluctuating energy densities (in the transverse
plane). Viscosity will quickly reduces these strong fluctuations. 

\begin{itemize}
\item We try to mimic viscous effects by taking artificially large values
of the flux tube radii (we take 1 fm), in order to get smoother initial
conditions.
\end{itemize}
This has the effect of reducing the elliptical flow by 20-30\%, as
needed.%
\begin{figure}[tb]
\begin{centering}
\includegraphics[scale=0.38]{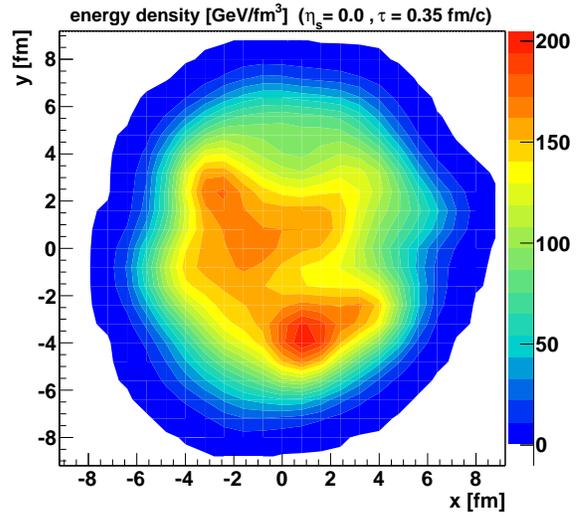}
\par\end{centering}

\caption{(Color online) Initial energy density in a central Pb-Pb collision
at 2.76 TeV, at a space-time rapidity $\eta_{s}=0$, as a function
of the transverse coordinates $x$ and y. \label{cap:eiau1} We take
artificially large values of the flux tube radii -- which provides
relatively smooth initial conditions -- to mimic viscous effects.}

\end{figure}
In Fig. \ref{cap:eiau1}, we show an example of such an initial energy
density.

\section{Centrality dependence of the multiplicity and fake scaling laws}

As a very first check of the new approach, we consider the centrality
dependence of the charged particle yield. Although very basic, there
is quite some confusion about this quantity. Whereas hard processes
scale roughly with the number of binary nucleon-nucleon collisions
(in a simple geometrical picture), the centrality dependence of the
charged particle yield (dominated by low $p_{t}$) is very different:
it looks more like a scaling with the number of participating nucleons.
This reminds us of the good old {}``wounded nucleon model'', which
has a physical meaning -- at low energies: the projectile and target
nucleons are excited ({}``wounded''), and this is the main source
of particle production. 

Amazingly, this approximate participant scaling holds also at higher
energies, the centrality dependence at the LHC is almost identical
to the one at the RHIC \citet{rapCMS}. This is quite strange, since
one might believe that at higher energies hard processes dominate,
so one could expect more binary scaling. But this is not the case.

\begin{figure}[h]
\begin{centering}
\includegraphics[angle=270,scale=0.26]{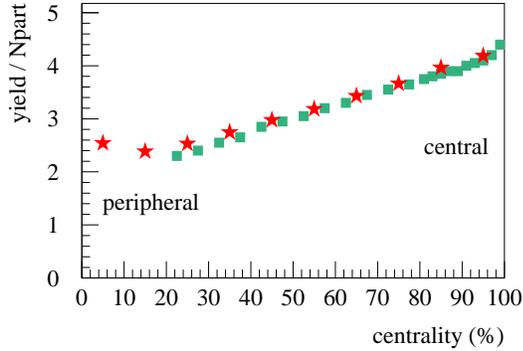}
\par\end{centering}

\caption{(Color online) Centrality dependence of the charged particle yield.
Calculations (stars) are compared to ATLAS data \citet{rapATLAS}
(circles). \label{cap:yield}}

\end{figure}

What do we get in the multiple scattering approach? In Fig. \ref{cap:yield},
we plot the yield per participant (~$dn/d\eta(0)$/Npart~) as a
function of the centrality. Npart is the number of participating nucleons.
As in the data \citet{rapATLAS}, we obtain a moderately increasing
yield per participant. How can this happen? How can one get something
like a wounded nucleon result at the LHC? In the model we can of course
easily check the relative contribution of particle production from
remnant decays.%
\begin{figure}[h]
\begin{centering}
\includegraphics[angle=270,scale=0.26]{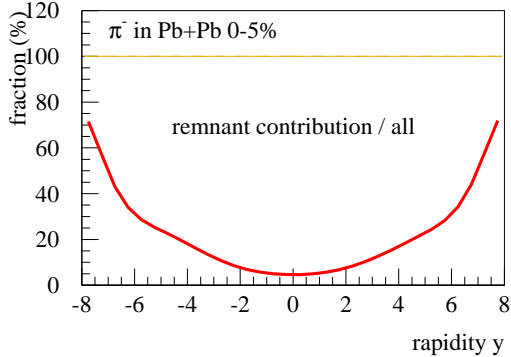}
\par\end{centering}

\caption{(Color online) Relative fraction of particle production from remnants
versus rapidity. The contribution from remnants at mid-rapidity is
very small. \label{cap:remnant}}

\end{figure}
In Fig. \ref{cap:remnant}, we plot the relative fraction of particle
production from remnants as a function of the rapidity. As expected,
remnant particle production is important at large rapidities, but
the contribution at mid-rapidity is close to zero. So the physical
mechanism of soft particle production is not a wounded nucleon picture.

\begin{figure}[t]
\begin{centering}
\includegraphics[scale=0.21]{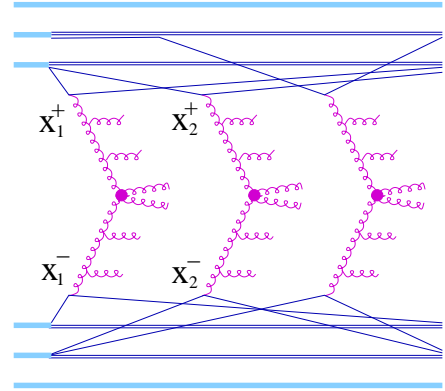}
\par\end{centering}

\caption{(Color online) Energy sharing in AA scatterings. The sum of all {}``ladder
end'' light cone momentum fractions $x_{k}^{\pm}$ linked to a given
remnant and the remnant fraction $x_{\mathrm{remn}\, i}^{\pm}$ have
to add up to unity. \label{ldmultaa2.eps}}

\end{figure}

In our approach, the source of particle production is the flux tubes,
originating from elementary scatterings, which are in principle proportional
to the number of binary nucleon-nucleon collisions. But there are
important effect due to energy conservation and shadowing, discussed
in detail in \citet{epos2,hajo}. In our multiple scattering approach
(which determines the initial conditions), the complete AA scattering
amplitude is expressed in terms of elementary contributions, which
are parton ladders, later showing up as strings. Each parton ladder
is characterized by the light cone momentum fractions $x_{k}^{+}$
and $x_{k}^{-}$ of the {}``ladder ends'', which are the outer partons
of the ladder, see Fig. \ref{ldmultaa2.eps} (also transverse momenta
are considered, but not discussed here). %

It is a unique feature of our approach that we do a precise bookkeeping
of energy and momentum: For each nucleon (projectile or target) the
initial energy-momentum has to be shared by all the ladders connected
to this nucleon and the nucleon remnants, i.e. for all nucleons $i$
we have\begin{equation}
\sum_{\begin{array}{c}
\mathrm{all\,\, ladders}\, k\,\\
\mathrm{connected\, to\, nucleon}\, i\end{array}}\!\!\!\!\!\! x_{k}^{\pm}\quad+\quad x_{\mathrm{remn}\, i}^{\pm}\quad=\quad1\,,\end{equation}
where $x_{\mathrm{remn}\, i}^{\pm}$ is the momentum fraction of the
nucleon remnant $i$. These are very strong conditions, which affect
the results substantially, see \citet{hajo}. 

The most important consequence relevant for our discussion here is
the fact that parton ladders leading to low $p_{t}$ particles are
suppressed compared to what is expected from binary scaling. %
\begin{figure}[tb]
\begin{centering}
\includegraphics[angle=270,scale=0.35]{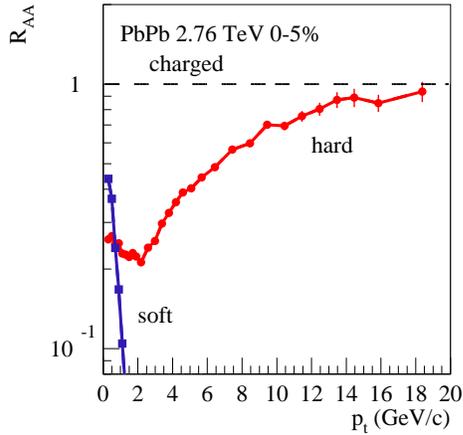}
\par\end{centering}

\caption{(Color online) Nuclear modification factor $R_{AA}=N_{\mathrm{coll}}^{-1}\,(dn_{AA}/dp_{t})\,/\,(dn_{pp}/dp_{t})$
vs transverse momentum $p_{t}$, showing the breaking of binary scaling
at low $p_{t}$ (due to energy conservation). The resulting {}``approximate
participant scaling at low $p_{t}$'' is a pure coincidence! \label{grt1.eps}}

\end{figure}
We will get a nuclear modification factor which is less than one at
low $p_{t}$, as shown in Fig. \ref{grt1.eps} for central Pb-Pb collisions
at 2.76 TeV. So although particle production at central rapidities
in very high energy collisions is dominated by binary scattering (providing
the initial energy density), particle production does not increase
proportional to the number of binary scatterings, due to energy conservation. 

It is absolutely necessary that binary scaling is broken at low $p_{t}$,
because it is simply an experimental fact. The usual explanation is
a two component picture: hard scattering at high $p_{t}$ which shows
binary scaling and a soft component which scales as the number of
participants In our picture, binary collisions determine everything.
But certain binary collisions are suppressed due to energy conservation,
leading to a deviation from $R_{AA}=1$.

\begin{figure}[tb]
\begin{centering}
\includegraphics[angle=270,scale=0.35]{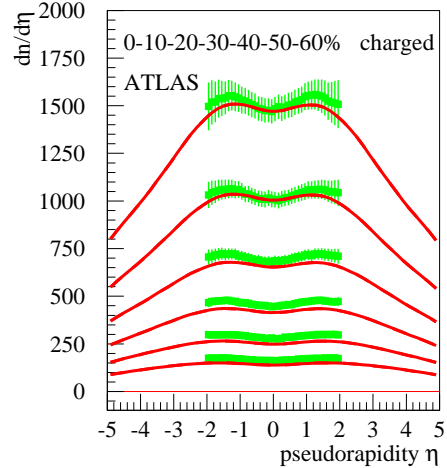}
\par\end{centering}

\caption{(Color online) Pseudorapidity distributions of charged particles for
different centralities. The lines are calculations, the the circles
are ATLAS data \citet{rapATLAS}, see text.\label{fig:rapidity} }

\end{figure}

We will discuss the $p_{t}$ dependence of $R_{AA}$ in the next section.
Here we present for completeness the pseudorapidity distributions
of charged particles for different centralities, see Fig. \ref{fig:rapidity},
where we compare our calculation with data from ATLAS \citet{rapATLAS}.%

\section{Transverse momentum dependence of particle yields: Importance of
hadronic rescattering of soft and jet hadrons}

\begin{figure}[tb]
\begin{centering}
\includegraphics[angle=270,scale=0.35]{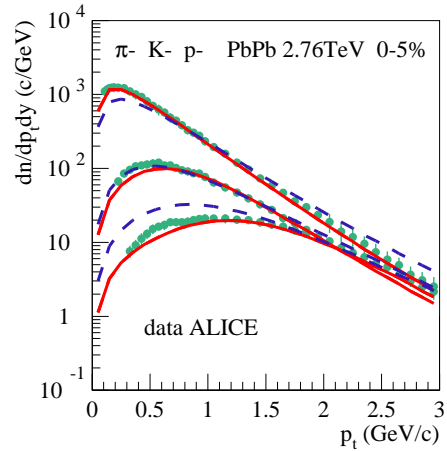}
\par\end{centering}

\caption{(Color online) Transverse momentum distributions of (from top to bottom)
negative pion, kaons, and protons, in the 0--5\% most central Pb-Pb
collisions at 2.76 TeV. We show the full calculation (solid lines)
and the ones without hadronic cascade (dashed lines), compared to
ALICE data (circles) \citet{ptALICE}.\label{fig:pt}}

\end{figure}

\begin{figure}[tb]
\begin{centering}
\includegraphics[angle=270,scale=0.35]{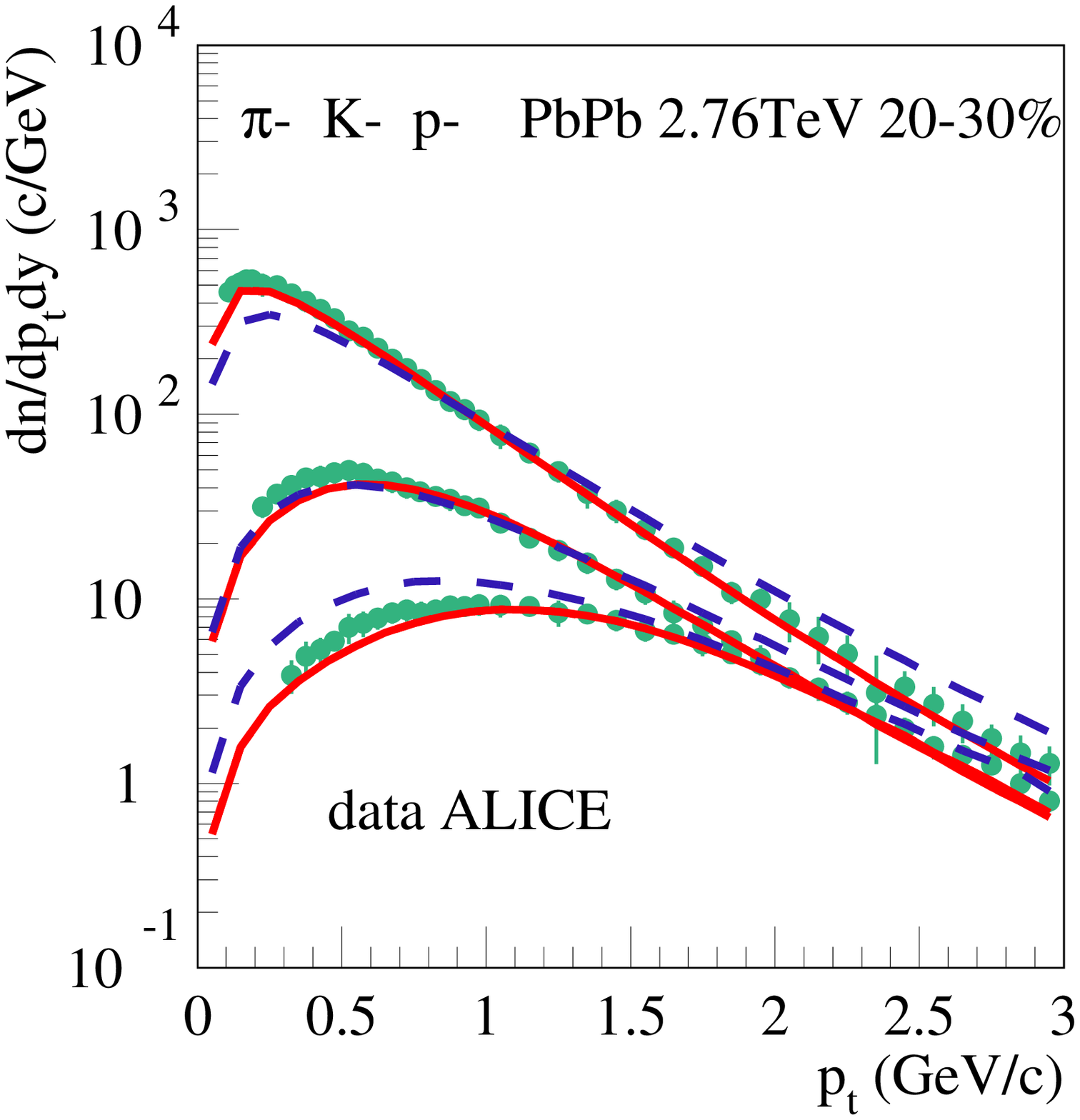}
\par\end{centering}

\caption{(Color online) Same as Fig. \ref{fig:pt}, but 20--30\% most central.\label{fig:pt2}}

\end{figure}

\begin{figure}[tb]
\begin{centering}
\includegraphics[angle=270,scale=0.35]{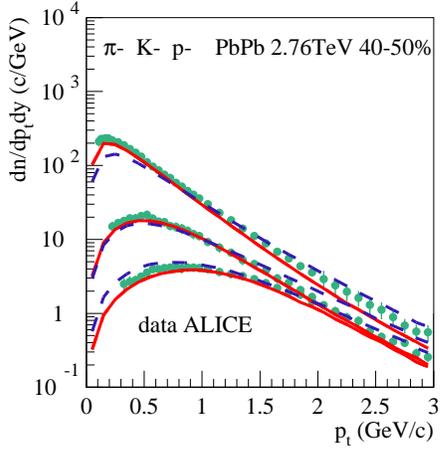}
\par\end{centering}

\caption{(Color online) Same as Fig. \ref{fig:pt}, but 40--50\% most central.\label{fig:pt3}}

\end{figure}

We first investigate particle production at low transverse momenta.
In figs. \ref{fig:pt}, \ref{fig:pt2}, and \ref{fig:pt3}, we show
transverse momentum distributions of pions, kaons, and protons, for
central and semi-peripheral Pb-Pb collisions at 2.76 TeV. We compare
the full calculation including hydrodynamic evolution and hadronic
final state cascade (solid lines) with the calculation without cascade
(dashed lines) and with data from ALICE \citet{ptALICE}.%

In order to understand the results, one has to recall that not only
the {}``soft'' particles produced from the fluid may interact, but
also the jet-particles having enough energy to escape the fluid may
interact with these soft particles. In particular intermediate $p_{t}$
jet particles are candidates, because their formation time will produce
them just in the high density hadronic region. Let us discuss the
consequences of these interactions, by comparing the solid and dashed
curves in the figures.

\begin{figure}[tb]
\begin{centering}
\includegraphics[angle=270,scale=0.35]{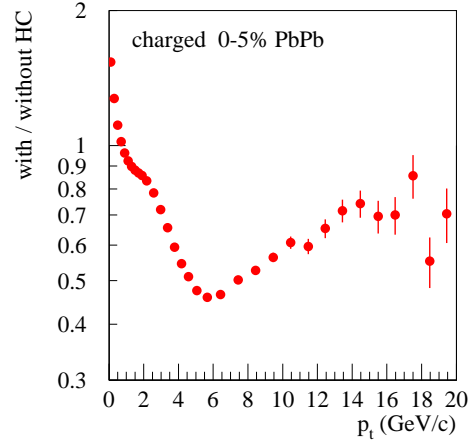}
\par\end{centering}

\caption{(Color online) Transverse momentum dependence of the ratio of the
full calculation to the one without hadronic cascade, for charged
particle production in the 0--5\% most central Pb-Pb collisions at
2.76 TeV.\label{fig:pt4}}

\end{figure}

\begin{itemize}
\item We see in particular in Fig. \ref{fig:pt} a strong reduction of protons
at low $p_{t}$ due to hadronic rescattering, which can be attributed
to proton-antiproton annihilation among the soft hadrons. 
\item We see also a sizable increase of pion production at low $p_{t}$,
which is due to inelastic rescatterings of jet hadrons with soft ones 
\end{itemize}
In figs. \ref{fig:pt}, \ref{fig:pt2}, and \ref{fig:pt3}, we only
show results up to 3 GeV/c, because this is the range where data on
protons, pions, and kaons are available. It is nevertheless interesting
to know the effect of jet-soft scattering beyond 3 GeV/c. We therefore
plot in Fig. \ref{fig:pt4} the ratio of the full calculation to the
one without hadronic cascade, for the $p_{t}$ spectra %
\begin{figure}[tb]
\begin{centering}
\includegraphics[angle=270,scale=0.35]{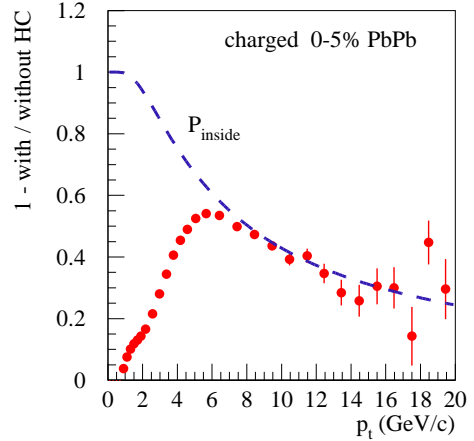}
\par\end{centering}

\caption{(Color online) Transverse momentum dependence of $1-R$ (red dots),
with $R$ being the ratio plotted in Fig. \ref{fig:pt4}. The dashed
line is the estimate $P_{\mathrm{inside}}$of the probability to produce
a jet hadron inside the fluid.\label{fig:pt5}}

\end{figure}
of charged particles (dominated by pions) in central Pb-Pb collisions
at 2.76 TeV, up to 20 GeV/c. There is a big effect at intermediate
values of $p_{t}$ -- up to 20~GeV/c~! In other words, jet-soft
rescattering is very important in this range. Similar observation
have already been made in \citet{carsten} for AuAu collisions at
the RHIC.

The big effect of the jet-soft interaction can be understood by plotting
$1-R,$ with $R$ being the ratio (with / without cascade) plotted
in Fig. \ref{fig:pt4}, together with the probability estimate $P_{\mathrm{inside}}$
to produce a jet (pre)hadron inside the fluid, see fig. \ref{fig:pt5}.
These early produced hadrons go through the dense hadronic phase (of
soft hadrons), and $P_{\mathrm{inside}}$ is therefore also a measure
of the probability of having a jet-soft interaction. We see indeed\begin{equation}
1-R=P_{\mathrm{inside}},\end{equation}
at large $p_{t}$ (in absolute terms, without adding factors). Even
though we are running out of statistics, it is clear from the above
discussion that the effect goes well beyond 20 GeV/c.

To compare the $p_{t}$ spectra with experimental data, one uses often
the so-called nuclear modification factor $R_{AA}$, which is the
ratio of the inclusive transverse momentum spectrum of particles in
nucleus-nucleus scatterings over the proton-proton ones, normalized
by the number $N_{\mathrm{coll}}$ of binary collisions. %
\begin{figure}[tb]
\begin{centering}
\includegraphics[angle=270,scale=0.35]{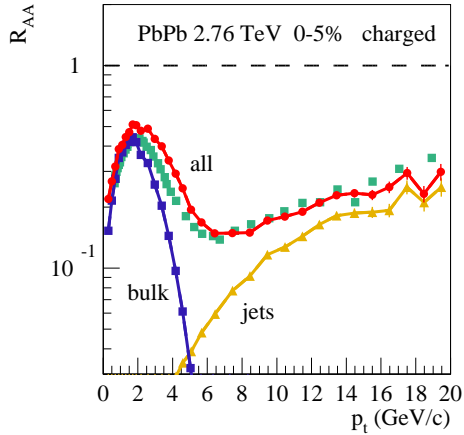}
\par\end{centering}

\caption{(Color online) The nuclear modification factor in Pb-Pb at 2.76 TeV
vs $p_{t}$: We compare data \citet{raaALICE} (squares) with the
full calculation (red line + circles) and its jet contribution (yellow
line + triangles) , as well as the bulk (hydro) contribution of a
calculation without hadronic cascade (blue line + squares) .\label{z-seljet-2.eps}}

\end{figure}
Doing this procedure, we obtain the curves shown in Fig. \ref{z-seljet-2.eps},
where we plot our simulation results for charged particle production
together with the data from ALICE \citet{raaALICE}. We show the full
model, including hydrodynamic evolution and final state hadronic cascade
\citet{urqmd} and its jet contribution from the string segments which
escaped from the bulk and which did not rescatter. We also show the
bulk contribution (originating from the hadronized fluid) from the
calculation without final state hadronic cascade. The two latter curves
do not add up to give the full result -- the difference is due to
the {}``secondary interactions'' discussed earlier:

\begin{itemize}
\item Fluid-jet interaction -- pushing the jet hadrons at intermediate $p_{t}$
to higher transverse momenta.
\item Jet-soft interactions between jet hadrons and soft ones from fluid
freeze-out.
\end{itemize}
There are also soft-soft interactions (among soft hadrons from fluid
freeze-out), which are important for baryon yields, but not so much
for the charged particle results.

\begin{figure}[tb]
\begin{centering}
\includegraphics[scale=0.19]{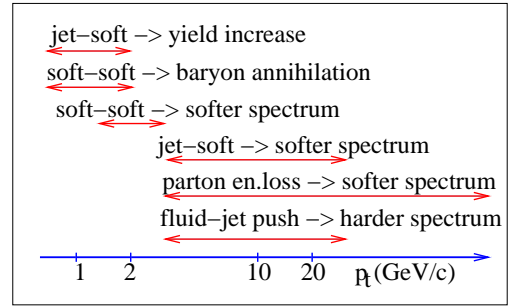}
\par\end{centering}

\caption{(Color online) Secondary interactions. The red arrows indicate the
$p_{t}$ range which are affected.\label{fig:Secondary}}

\end{figure}

From the above discussion it is clear that even considering elementary
quantities as charged particle yields, it is difficult to make any
quantitative analysis without considering these {}``secondary interactions''.
We sketch the different interactions in Fig. \ref{fig:Secondary}.

\section{Dihadron correlations in Pb-Pb at 2.76 TeV}

Our prescription for bulk-jet separation and interaction should also
strongly affect dihadron correlations, which provide much more information
than simple spectra. With all parameters ($k_{\mathrm{Eloss}}$, $E_{0}$,
$\tau_{\mathrm{seg}}$, $p_{\mathrm{diq}}$) being fixed from the
considerations in the last section, we now compute dihadron correlation
functions defined as \begin{equation}
R(\Delta\eta,\Delta\phi)=\frac{M}{S}\times\frac{S(\Delta\eta,\Delta\phi)}{M(\Delta\eta,\Delta\phi)},\end{equation}
where $S$ is the number of pairs in real events, and $M$ the number
of pairs for mixed events. 

\begin{figure}[b]
\begin{centering}
\includegraphics[scale=0.4]{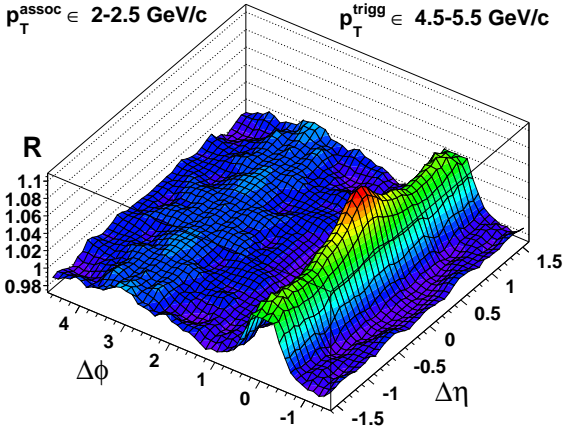}
\par\end{centering}

\caption{(Color online) Calculated dihadron correlation function for $p_{t}^{\mathrm{trigg}}$
in the interval 4.5-5.5 GeV/c and $p_{t}^{\mathrm{assoc}}$ in the
range 2-2.5 GeV/c in the 0--10\% most central Pb-Pb collisions at
2.76 TeV. \label{Q1}}

\end{figure}

As an example, we show in Fig. \ref{Q1} %
the correlation function for the $p_{t}$ of the trigger particle
($p_{t}^{\mathrm{trigg}}$) in the interval 4.5-5.5 GeV/c and the
$p_{t}$ of the associated particle ($p_{t}^{\mathrm{assoc}}$) in
the range 2-2.5 GeV/c, in the 0--10\% most central Pb-Pb collisions
at 2.76 TeV. Besides the jet peak at $\Delta\phi=0$ and $\Delta\eta=0$,
we clearly identify a completely flat ridge over the full range in
$\Delta\eta$ at $\Delta\phi=0$. 

The reason for the ridge structure is the fact that there is an azimuthal
asymmetry of the initial energy density (see Fig. \ref{cap:eiau1}).
Although the energy density is biggest around space-time rapidity
$\eta_{s}=0$ and drops fast towards forward and backward $\eta_{s}$,
the shape of the asymmetry is preserved. This leads finally to an
asymmetric flow, again very similar at different values of $\eta_{s}$,
and this {}``makes'' the long range correlation at $\Delta\phi=0$.

\begin{figure}[tb]
\begin{centering}
\includegraphics[scale=0.37]{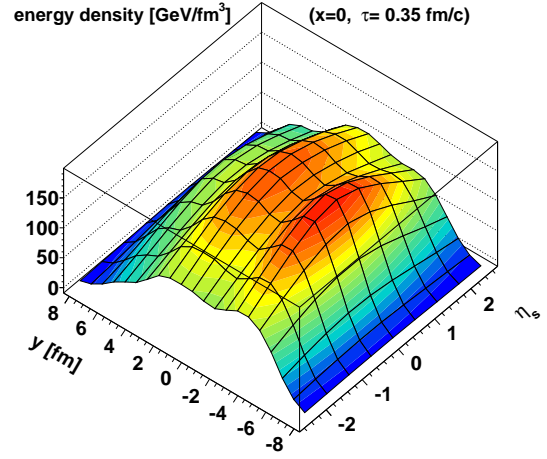}
\par\end{centering}

\caption{(Color online) The energy density for a single event in a central
Pb-Pb collisions at 2.76 TeV as a function of the longitudinal variable
$\eta_{s}$ and the transverse one $y$ (and for $x=0$). \label{edensity}}

\end{figure}

The smooth $\eta_{s}$ dependence of the energy density in our approach
(see Fig. \ref{edensity}) %
is due to the fact the energy density is calculated from flux tubes.
And these flux tubes have to be treated correctly as continuous longitudinal
objects (as we do). In an earlier version, we treated flux tubes via
the randomly (in $\eta_{s}$) distributed flux tube segments, obtained
from a string fragmentation procedures. This gives a bumpy structure
in $\eta_{s}$-- the ridge is not flat any more but has a Gaussian
shape! So the flux-tube basis is an essential ingredient for obtaining
a perfect ridge shape, as observed in the data.

\begin{figure}[tb]
\begin{centering}
\includegraphics[scale=0.4]{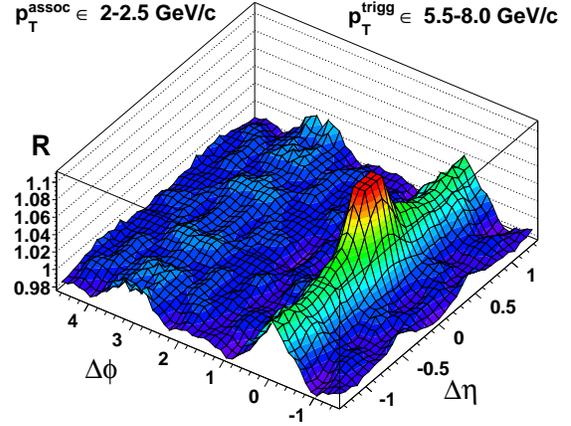}
\par\end{centering}

\caption{(Color online) Calculated dihadron correlation function for $p_{t}^{\mathrm{trigg}}$
in the interval 5.5-8.0 GeV/c and $p_{t}^{\mathrm{assoc}}$ in the
range 2-2.5 GeV/c in the 0--10\% most central Pb-Pb collisions at
2.76 TeV . \label{Q2}}

\end{figure}

\begin{figure}[tb]
\begin{centering}
\includegraphics[scale=0.28]{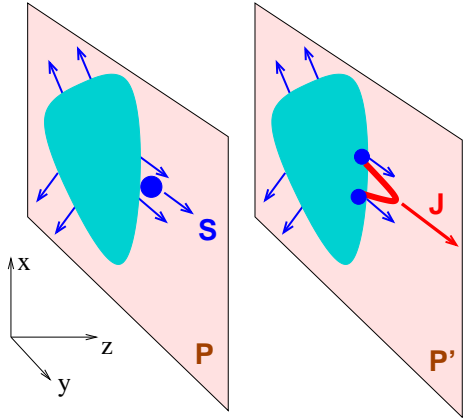}
\par\end{centering}

\caption{(Color online) Sketch of two cuts of the fluid volume corresponding
to the space-time rapidities $\eta_{s}$ and $\eta_{s}'$, the two
corresponding transverse planes being $\mathbf{P}$ and $\mathbf{P'}$.
We show the example of a triangular flow pattern -- the same at $\eta_{s}$
and $\eta_{s}'$. \label{ridge3}}

\end{figure}

In Fig. \ref{Q2}, %
we show a correlation function for $p_{t}^{\mathrm{trigg}}$ in the
interval 5.5-8.0 GeV/c and $p_{t}^{\mathrm{assoc}}$ in the range
2-2.5 GeV/c, again in the 0--10\% most central Pb-Pb collisions at
2.76 TeV. Although the trigger $p_{t}$ is too large to originate
from freeze-out (from the flowing fluid), one still observes a ridge
structure, which is due to the fluid-jet interaction. Let us consider
again the situation of an initial azimuthal anisotropy in the energy
density which is transported into a corresponding anisotropy in the
flow, as discussed earlier. We sketch in Fig. \ref{ridge3}%
the (somewhat exaggerated) situation of a triangular transverse flow
pattern with maximal flow around $\phi=0^{o}$, $120^{o}$, and $240^{o}$
(with respect to the $y$-axis). The flow maxima are indicated by
blue arrows. Again it is very important that this flow pattern is
(not necessarily in magnitude, but in shape) very similar at different
longitudinal positions -- in the figure indicated by the two transverse
planes $\mathbf{P}$ and $\mathbf{P'}$, corresponding to two different
space-time rapidities $\eta_{s}$ and $\eta_{s}'$. A soft hadron
($\mathbf{S}$) produced at $\eta_{s}$ at the fluid surface close
to the position of maximal flow (for example at $\phi=0^{o}$), will
be boosted by the latter one and therefore carry information about
this flow. A jet hadron ($\mathbf{J}$) produced at $\eta_{s}'$ at
the same angle ($\phi=0^{o}$) close to the surface, will pick up
a quark and an antiquark, both carrying flow, which adds the corresponding
transverse momentum to the $p_{t}$ of the string segment (red element
in the figure). It is the same flow which affects the jet hadron at
$\eta_{s}'$ and the soft hadron at $\eta_{s}$, which creates the
dihadron correlation at $\Delta\phi=0$, the {}``ridge''. The correlation
remains visible, even when the flow contribution to the jet hadron
is only 10\%, this is why the correlation is still present even for
trigger transverse momenta beyond 10 GeV/c.

\begin{figure}[tb]
\begin{centering}
\includegraphics[scale=0.4]{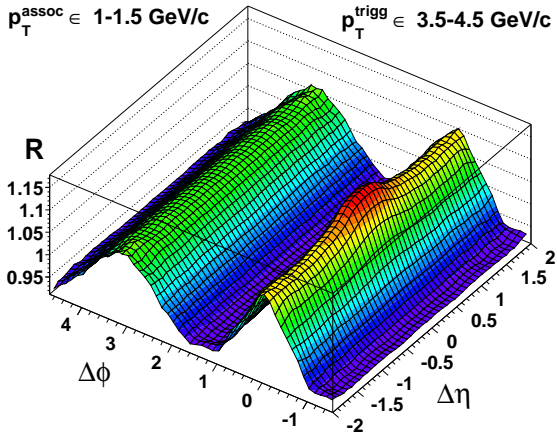}
\par\end{centering}

\caption{(Color online) Calculated dihadron correlation function for $p_{t}^{\mathrm{trigg}}$
in the interval 3.5-4.5 GeV/c and $p_{t}^{\mathrm{assoc}}$ in the
range 1-1.5 GeV/c in the 40--50\% most central Pb-Pb collisions at
2.76 TeV. \label{Q3}}

\end{figure}

\begin{figure}[tb]
\begin{centering}
\includegraphics[scale=0.4]{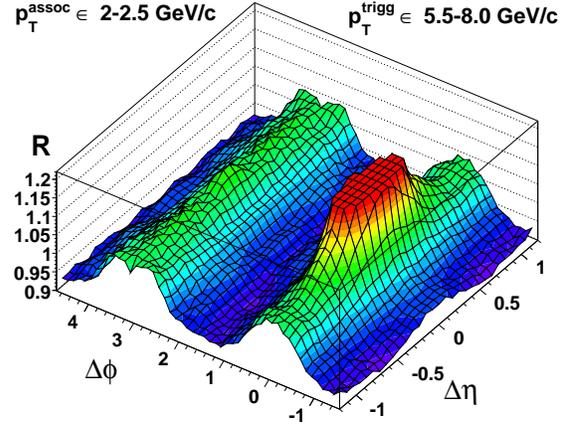}
\par\end{centering}

\caption{(Color online) Calculated dihadron correlation function for $p_{t}^{\mathrm{trigg}}$
in the interval 5.5-8.0 GeV/c and $p_{t}^{\mathrm{assoc}}$ in the
range 2-2.5 GeV/c in the 40--50\% most central Pb-Pb collisions at
2.76 TeV. The jet peak has been cut for better visibility. \label{Q4}}

\end{figure}

We will now discuss some examples of semi-peripheral Pb-Pb collisions
at 2.76 GeV/c. In Fig. \ref{Q3}, %
we show the correlation function for $p_{t}^{\mathrm{trigg}}$ in
the interval 3.5-4.5 GeV/c and $p_{t}^{\mathrm{assoc}}$ in the range
1-1.5 GeV/c, in the 40--50\% most central Pb-Pb collisions at 2.76
TeV. It can be clearly seen from the figure that the elliptical flow
($\sim\cos(2\Delta\phi$)) is dominant, besides the jet peak at $\Delta\eta=0,\,\Delta\phi=0$.
But also here higher order harmonics ($\sim\cos(i\Delta\phi$)) contribute,
as we will discuss later. In Fig. \ref{Q4},%
we show another example of a correlation function for semi-peripheral
collisions, with somewhat bigger trigger $p_{t}$. We use $p_{t}^{\mathrm{trigg}}$
in the interval 5.5-8.0 GeV/c and $p_{t}^{\mathrm{assoc}}$ in the
range 2-2.5 GeV/c. The $\Delta\eta$ range is chosen smaller to avoid
to big statistical fluctuations. The jet contribution becomes dominant
in this case, but when we cut off the jet peak, we clearly see a very
similar elliptical flow structure as in the previous example.

The correlation functions are essentially flat as a function of $\Delta\eta$,
for large $\Delta\eta$. One therefore gets complete information about
the long range correlations by integrating over $\Delta\eta$,\begin{equation}
R(\Delta\phi)=\frac{1}{2(B-A)}\int_{A<|\Delta\eta|<B}R(\Delta\eta,\Delta\phi)\, d\Delta\eta,\end{equation}
where we use $A=0.8$ and the maximum $B=2$. This function agrees
perfectly with its Fourier decomposition, \begin{equation}
R(\Delta\phi)=1+\sum_{n=1}^{5}2V_{n\Delta}\cos(n\Delta\phi),\end{equation}
using the first five terms. This is very convenient, because it allows
to discuss the features of the correlation functions for different
options for $p_{t}^{\mathrm{trigg}}$ and $p_{t}^{\mathrm{assoc}}$
by simply considering the Fourier coefficients. 

\begin{figure}[tb]
\begin{raggedright}
{\large (a)}
\par\end{raggedright}{\large \par}

\begin{raggedright}
\vspace*{-1.cm}
\par\end{raggedright}

\begin{centering}
$\qquad$\includegraphics[angle=270,scale=0.31]{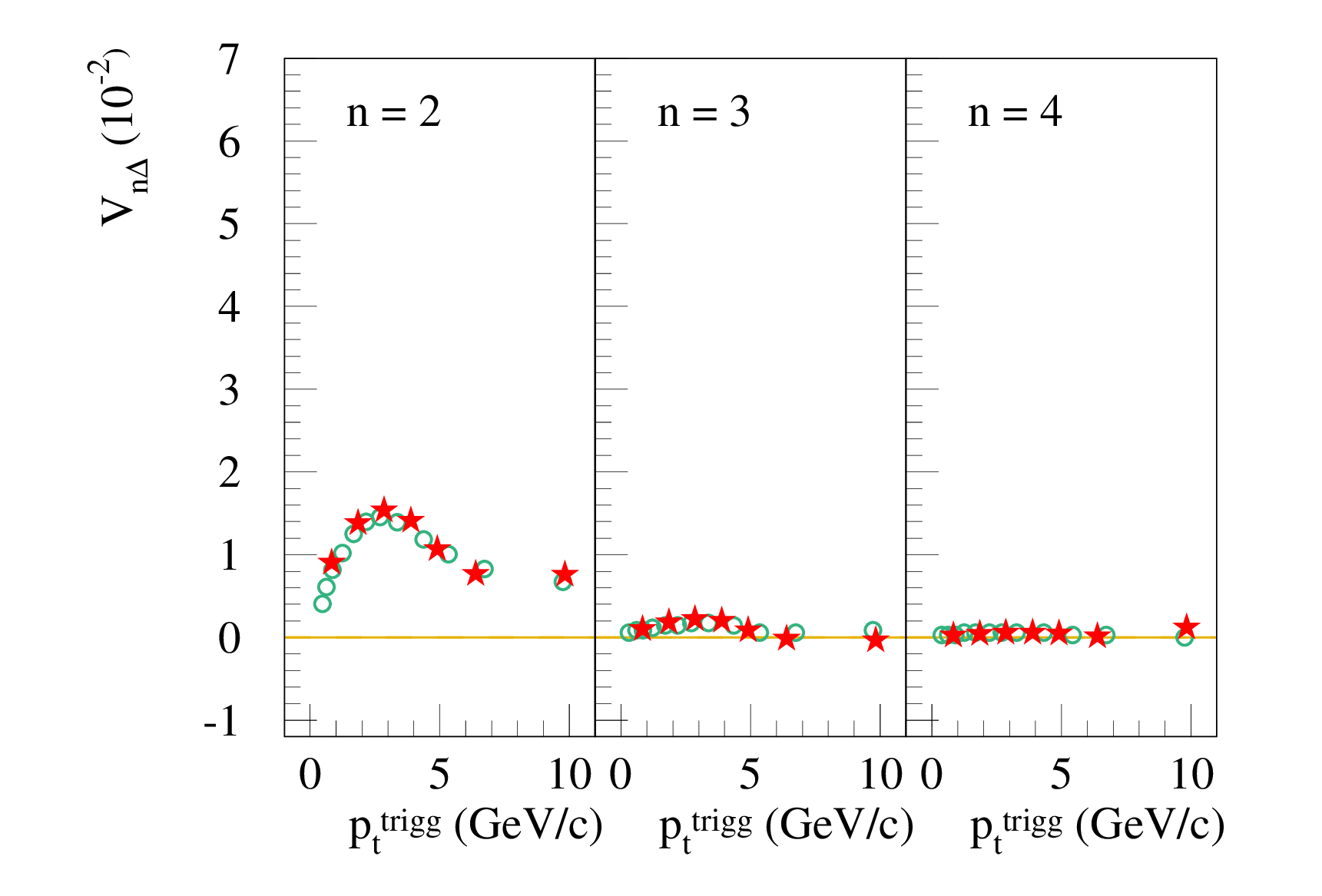}
\par\end{centering}

\begin{raggedright}
{\large (b)}
\par\end{raggedright}{\large \par}

\begin{raggedright}
\vspace*{-1.cm}
\par\end{raggedright}

\begin{centering}
$\qquad$\includegraphics[angle=270,scale=0.31]{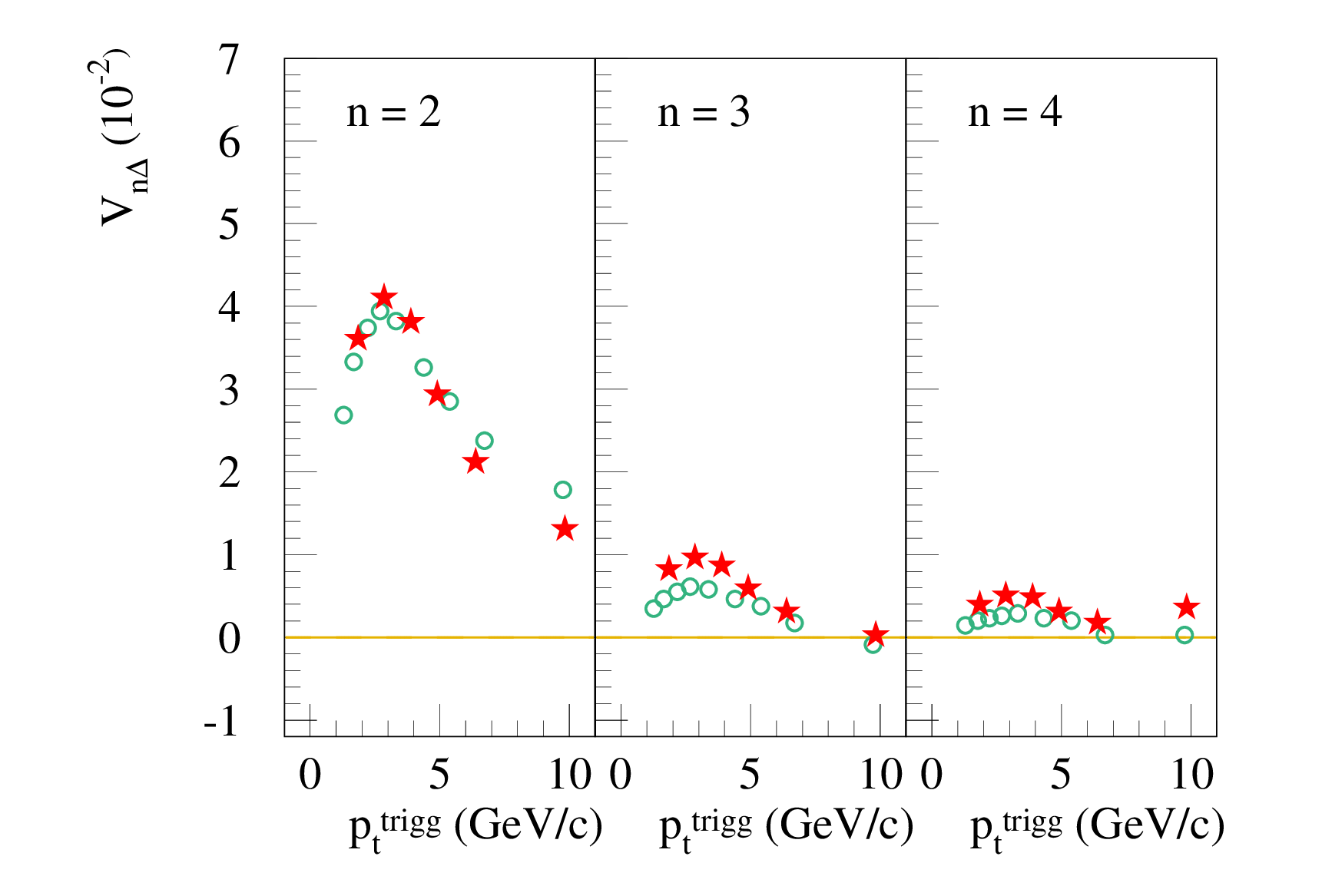}
\par\end{centering}

\begin{raggedright}
{\large (c)}
\par\end{raggedright}{\large \par}

\begin{raggedright}
\vspace*{-1.cm}
\par\end{raggedright}

\begin{centering}
$\qquad$\includegraphics[angle=270,scale=0.31]{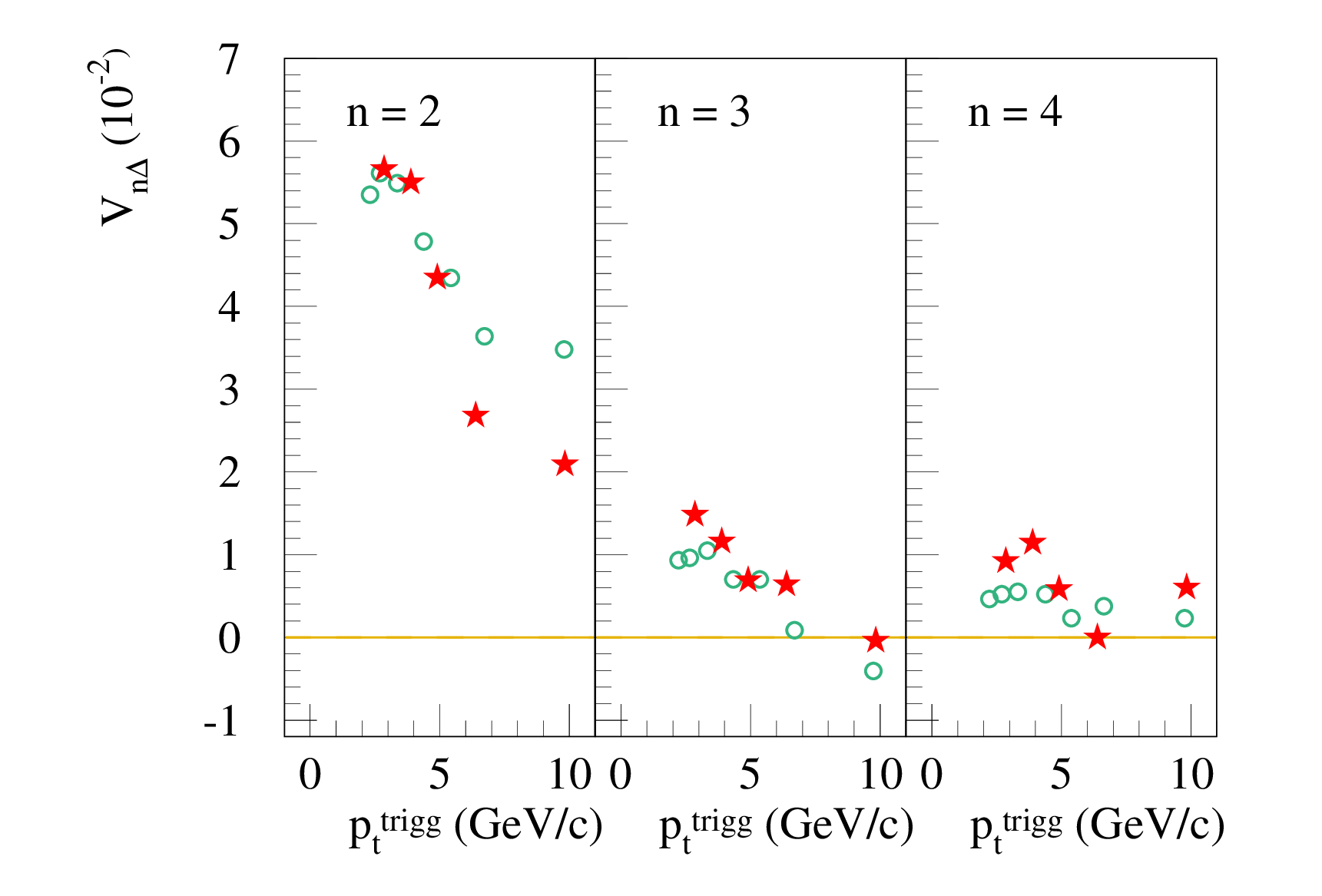}
\par\end{centering}

\caption{(Color online) The Fourier coefficients $V_{n\Delta}$ as a function
of $p_{t}^{\mathrm{trigg}}$ for $p_{t}^{\mathrm{assoc}}$ within
0.25-0.5~GeV/c (a), 1-1.5~GeV/c (b), 2-2.5~GeV/c (c), in the 40-50\%
most central Pb-Pb collisions at 2.76 TeV. We compare the ALICE data
\citet{ridgeALI} (circles) with calculations (red stars) .\label{fou1}}

\end{figure}

\begin{figure}[tb]
\begin{raggedright}
{\large (a)}
\par\end{raggedright}{\large \par}

\begin{raggedright}
\vspace*{-1.cm}
\par\end{raggedright}

\begin{centering}
$\qquad$\includegraphics[angle=270,scale=0.31]{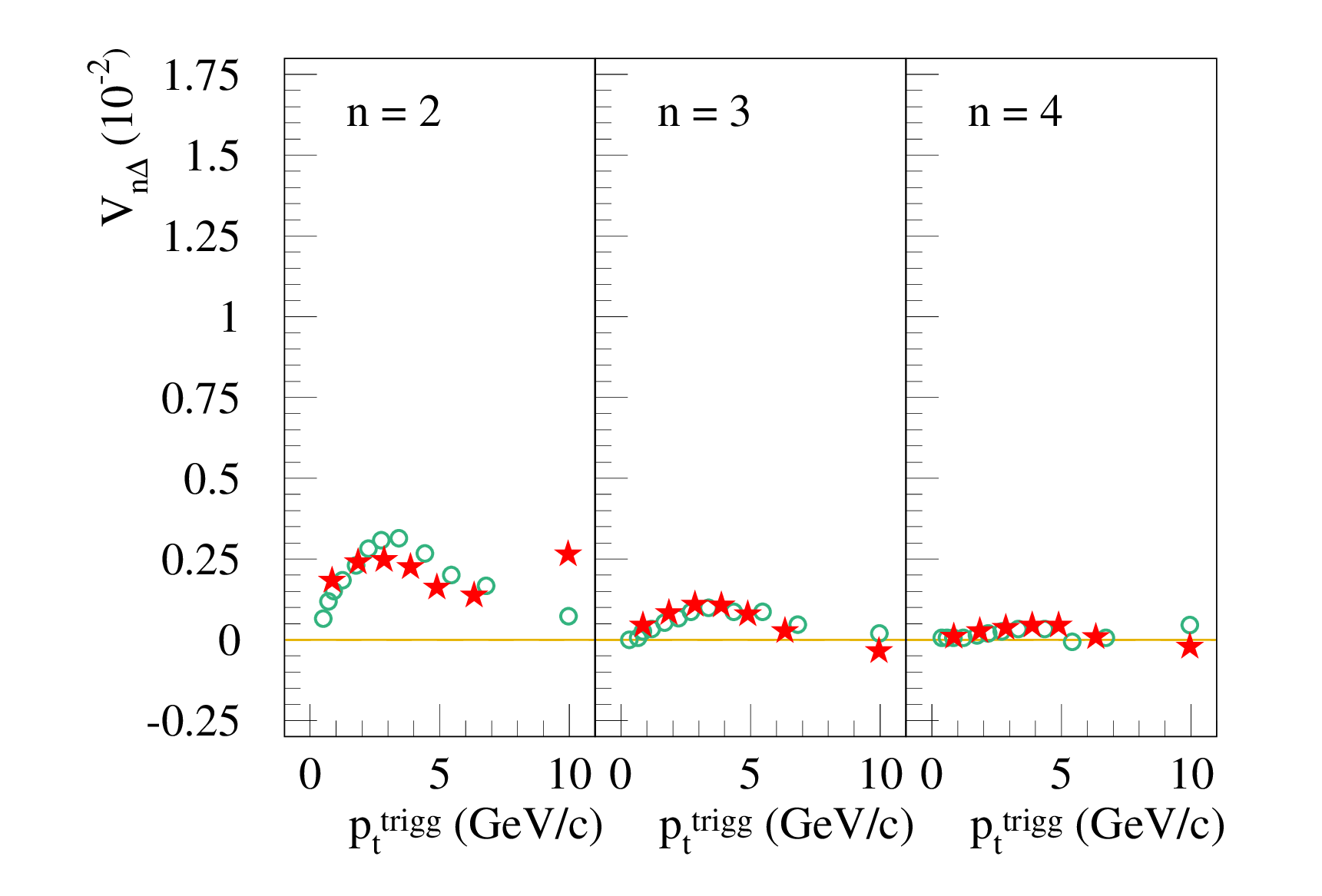}
\par\end{centering}

\begin{raggedright}
{\large (b)}
\par\end{raggedright}{\large \par}

\begin{raggedright}
\vspace*{-1.cm}
\par\end{raggedright}

\begin{centering}
$\qquad$\includegraphics[angle=270,scale=0.31]{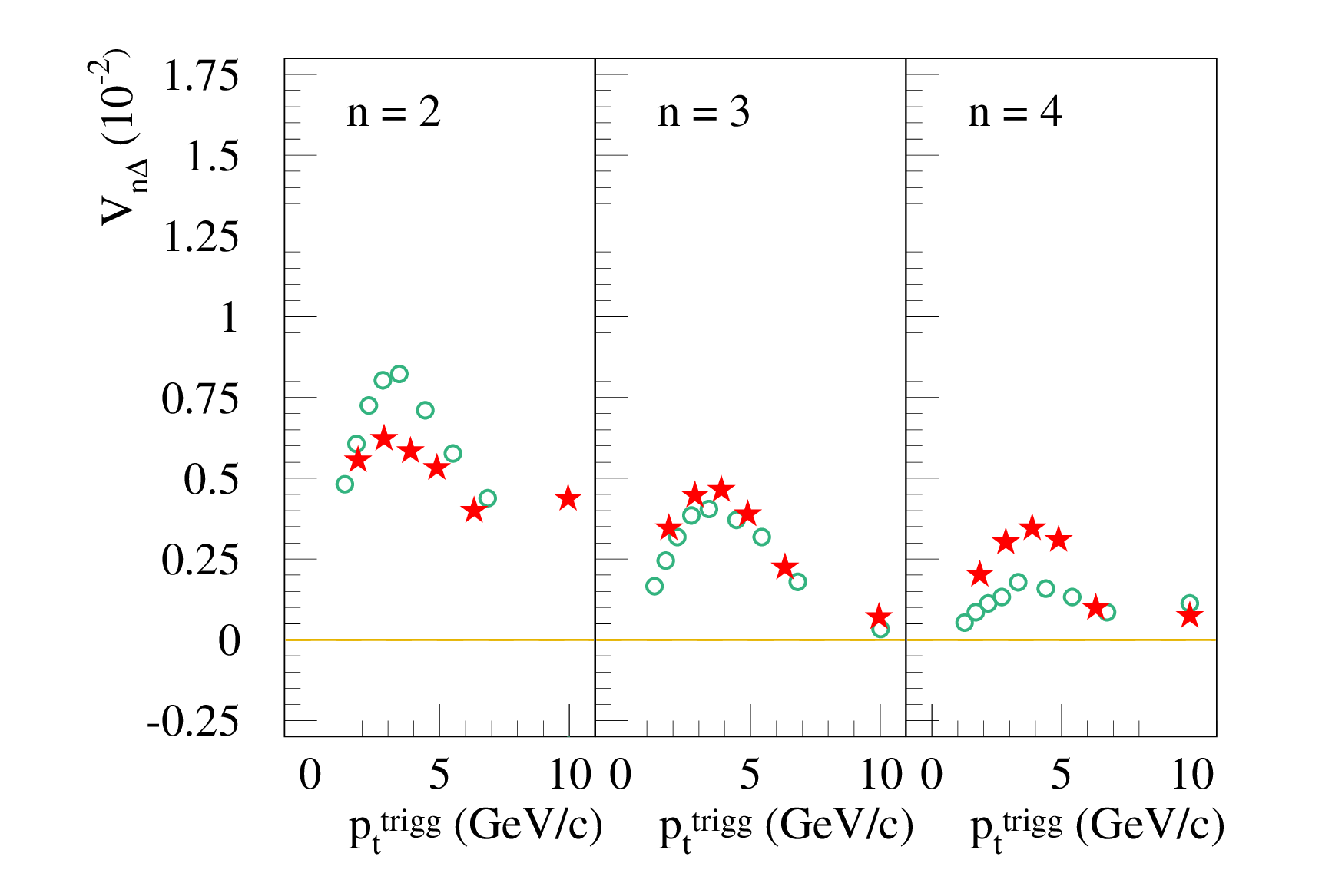}
\par\end{centering}

\begin{raggedright}
{\large (c)}
\par\end{raggedright}{\large \par}

\begin{raggedright}
\vspace*{-1.cm}
\par\end{raggedright}

\begin{centering}
$\qquad$\includegraphics[angle=270,scale=0.31]{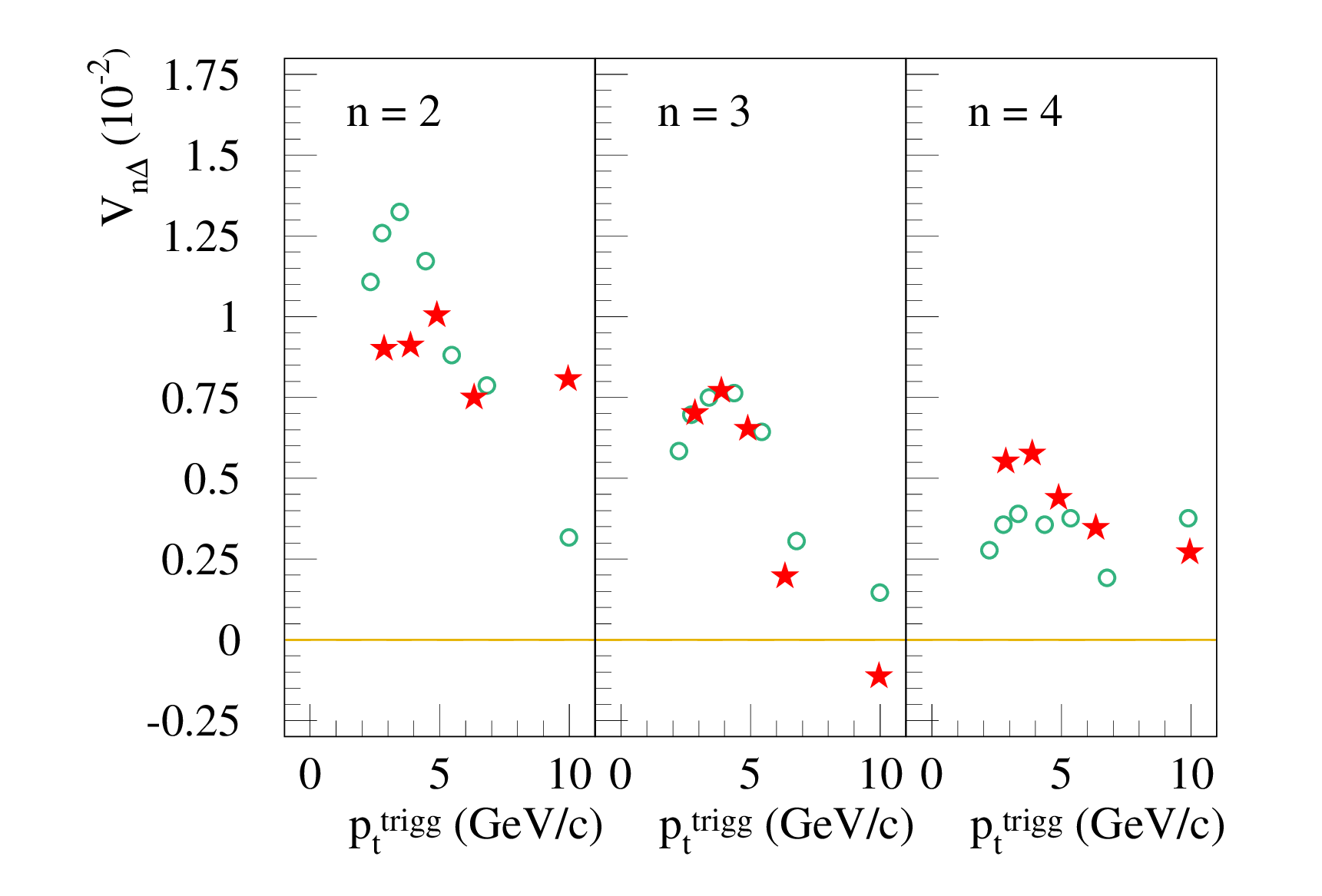}
\par\end{centering}

\caption{(Color online) Same as Fig. \ref{fou1}, but for the 0-10\% most central
Pb-Pb collisions at 2.76 TeV. We compare the ALICE data \citet{ridgeALI}
(circles) with calculations (red stars) .\label{fou2}}

\end{figure}

In figs. \ref{fou1} and \ref{fou2}, we plot some coefficients $V_{n\Delta}$
as a function of $p_{t}^{\mathrm{trigg}}$ for different intervals
of $p_{t}^{\mathrm{assoc}}$. The value of $p_{t}^{\mathrm{trigg}}$
is actually the mean value in a certain interval, the largest interval
being 8-15~GeV/c.%
We compare our simulation (stars) with the results from ALICE \citet{ridgeALI}
(circles). 

In the semi-peripheral collisions of Fig. \ref{fou1}, we see clearly
the dominance of elliptical flow: the $n=2$ coefficients are by far
the largest. Nevertheless, also the higher harmonics contribute. We
see in all cases an increase of the coefficients with $p_{t}^{\mathrm{assoc}}$
and with $p_{t}^{\mathrm{trigg}}$ up to values of around 2-3~GeV/c.
At the latter values the hydrodynamic flow contributes the most to
the correlation between soft hadrons from the fluid. 

For higher transverse momenta, the coefficients get smaller, because
the correlation between soft particles  dies out. But $V_{2\Delta}$does
not at all drop to zero at high $p_{t}$ because here the correlations
between soft and jet particles come into play -- the jet particles
which suffered a push by the fluid, as discussed earlier (fluid-jet
interaction). The fluid transfers at maximum few GeV/c of transverse
momentum to the jet, but this is easily visible in the correlation
(even at 20 GeV/c).

The results for semi-peripheral collisions are very robust and depend
little on model parameters. The most important ingredient is the elliptical
initial shape on the energy density, given by the nuclear geometry.
The effects depend of course on the flow velocity at freeze-out, but
this is not a parameter but itself a robust result (with a maximum
around 0.7c). Finally the results depend on the jet formation time,
which should be around 1 fm/c (the value we actually took without
really attempting a fine-tuning). 

The $V_{2\Delta}$ coefficients for central Pb-Pb collisions (Fig.
\ref{fou2}) are first of all smaller as compared to the semi-peripheral
ones (note the different scales in figs. \ref{fou1} and \ref{fou2}),
simply because the dominant effect of large initial ellipticity from
the geometry is absent. Here the ellipticity is purely random. Apart
from this, we observe the same features as for the semi-peripheral
collisions: increase with transverse momenta up to 2-3 GeV/c, then
decrease. A big difference in central Pb-Pb collisions compared to
semi-peripheral ones is the fact that the higher harmonics and in
particular $V_{3\Delta}$ contribute substantially, because here both
elliptical and triangular initial shape are of random origin (and
therefore comparable), whereas for more peripheral collisions the
geometrical elliptical shape dominates everything else. 

\begin{figure}[tb]
\begin{centering}
\includegraphics[angle=270,scale=0.3]{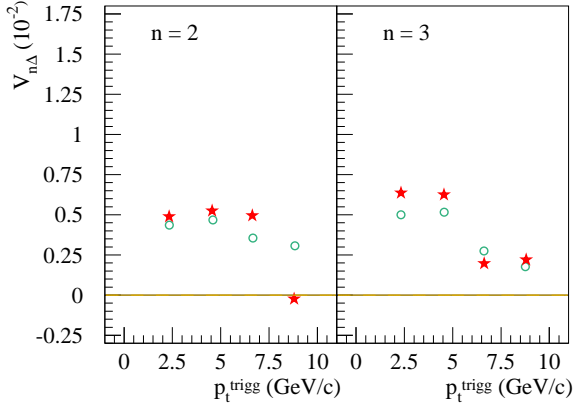}
\par\end{centering}

\caption{(Color online) The Fourier coefficients $V_{n\Delta}$ as a function
of $p_{t}^{\mathrm{trigg}}$ for $p_{t}^{\mathrm{assoc}}$ within
2-4 GeV/c in the 0-5\% most central Pb-Pb collisions at 2.76 TeV.
We compare the data \citet{ridgeCMS} (squares) with calculations
(red stars).\label{fou3}}

\end{figure}

In Fig. \ref{fou2} it seems that our calculation underestimates $V_{2\Delta}$,
in particular for the largest $p_{t}^{\mathrm{assoc}}$ range (2-2.5.~GeV).
Fortunately, similar data exist from CMS \citet{ridgeCMS}, for $p_{t}^{\mathrm{assoc}}$
in the range 2-4 GeV/c in the 0-5\% most central Pb-Pb collisions.
In Fig. \ref{fou3}, %
we plot the corresponding coefficients $V_{2\Delta}$ and $V_{3\Delta}$.
Here we slightly overpredict the data!

\section{v2 and formation times}

\begin{figure}[tb]
\begin{centering}
\hspace*{-0.7cm}\includegraphics[angle=270,scale=0.66]{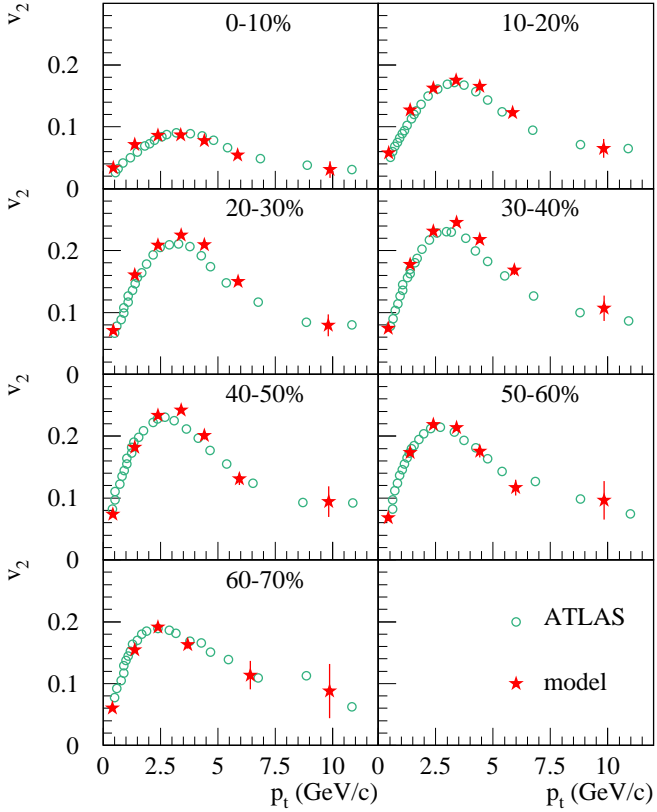}
\par\end{centering}

\caption{(Color online) $p_{t}$ dependence of elliptical flow (defined with
respect to the opposite hemisphere sub-event plane) for different
centralities in Pb-Pb collisions at 2.76 TeV. We compare the ATLAS
data \citet{v2ATLAS} (circles) with calculations (red lines).\label{v2}}

\end{figure}

Whereas dihadron correlations provide the most complete information
about particle production -- in particular concerning the role of
the {}``flowing{}`` fluid, one may get the essential information
by considering the elliptical flow coefficient $v_{2}$ of single
particle production, which is defined as \begin{equation}
v_{2}=\left\langle \cos[2(\phi-\phi_{\mathrm{Ref}})]\right\rangle ,\end{equation}
where $\phi$ is the azimuth angle of a particle, and $\phi_{\mathrm{Ref}}$
some reference plane. In \citet{v2ATLAS}, for the particles in the
forward (backward) $\eta$ hemisphere, the reference plane is the
event plane angle $\phi^{\mathrm{backward}}$ ($\phi^{\mathrm{forward}}$),
obtained from counting all particles in the opposite hemisphere. The
angles are obtained from \begin{equation}
\phi^{\mathrm{backward/\mathrm{forward}}}=\frac{1}{2}\tan^{-1}\frac{\left\langle \sin2\phi\right\rangle }{\left\langle \cos2\phi\right\rangle },\end{equation}
where the average is done in the forward / backward $\eta$ hemisphere
within $3.2<|\eta|<4.8$. The $v_{2}$ coefficient is then computed
as\begin{equation}
v_{2}=\left\langle \cos\left[2(\phi-\phi^{\mathrm{forward}/\mathrm{backward}})\right]\right\rangle .\end{equation}
Resolution correction is taken care of by dividing this expression
by $R=\sqrt{\left\langle \cos\left[2\Delta\phi\right]\right\rangle }$,
with $\Delta\phi=\phi^{\mathrm{backward}}-\phi^{\mathrm{forward}}$,
as in ref \citet{v2ATLAS}. Relating particles with the event plane
of the opposite hemisphere, we have kind of a long range correlation,
but less clean than using dihadron correlations with a $\Delta\eta>A$
requirement. But as mentioned before, the essential features can be
seen as well. In Fig. \ref{v2}, %
we plot $v_{2}$ as a function of the transverse momentum for different
centralities in Pb-Pb collisions at 2.76 TeV. The magnitude of the
elliptical flow coefficients increase at low $p_{t}$ to reach a maximum
around 2-3.5 GeV/c and then drop slowly at large $p_{t}$. 

\begin{figure}[tb]
\begin{centering}
\includegraphics[angle=270,scale=0.35]{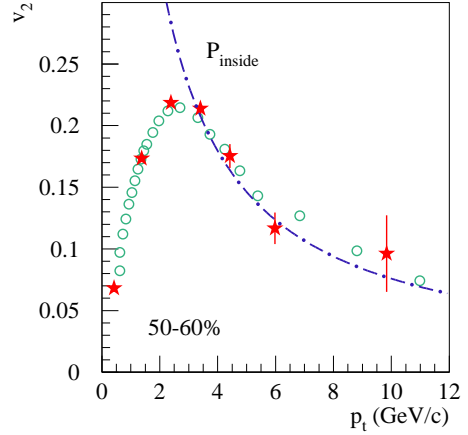}
\par\end{centering}

\caption{(Color online) Same data and model calculation as shown in Fig. \ref{v2}
for the case of 50-60\% most central collisions, together with $P_{\mathrm{inside}}$.\label{fig:estim}}

\end{figure}

The behavior at high $p_{t}$ is the most interesting aspect: even
at 10 GeV/c, there is a significant amount of elliptical flow, due
to the fluid-jet interaction, which pushes jet particles in the direction
of the collective flow at the freeze-out surface (and this effect
will continue up to even higher $p_{t}$, but we are simply running
out of statistics). The high $p_{t}$ behavior is closely related
to the formation time discussion we had earlier. The non-vanishing
$v_{2}$ at high $p_{t}$ is mainly due to fluid-jet interactions,
so the values should be related to the estimated probability $P_{\mathrm{inside}}$
to form the jet hadron inside the fluid, which is equivalent to the
fluid-jet interaction probability. In Fig. \ref{fig:estim}, we %
show $P_{\mathrm{inside}}$ (multiplied by an arbitrary factor), together
with the calculated and experimental $v_{2}$ already show in Fig.
\ref{v2}, which shows that $v_{2}$ is indeed proportional to the
fluid-jet interaction probability. To compute $P_{\mathrm{inside}}$
according to eq. (\ref{eq:iproba}), we use $c\tau_{\mathrm{form}}=1\,\mathrm{fm}$,
$mc^{2}=1\,\mathrm{GeV}$, $r_{\mathrm{Pb}}=6.5\,\mathrm{fm}$, and
\textbf{$b=11.5\,\mathrm{fm}$} (50-60\%). 

Having a finite $P_{\mathrm{inside}}$ is a necessary condition to
get a substantial $v_{2}$ at some large value of $p_{t}$, but not
a sufficient one. Let us therefore estimate the effect of the fluid-jet
interaction on $v_{2}$. One may consider a toy model, with soft particle
emission due to flow into some preferred azimuthal direction, say
$\phi_{\mathrm{flow}}$. Let us assume a jet hadron getting pushed
by the fluid in the direction of $\phi_{\mathrm{flow}}$, which corresponds
to adding some $\vec{p}_{t}^{\mathrm{\: soft}}$ to the {}``hard''
transverse momentum $\vec{p}_{t}^{\mathrm{\: hard}}$ of the flux-tube
segment. Without loss of generality, we may set $\phi_{\mathrm{flow}}=0$.
The total transverse momentum of the jet hadron is\begin{equation}
\vec{p}_{t}^{\mathrm{\: soft}}+\vec{p}_{t}^{\:\mathrm{hard}}=\left(\begin{array}{c}
p_{t}^{\mathrm{soft}}+p_{t}^{\mathrm{hard}}\cos(\phi)\\
p_{t}^{\mathrm{hard}}\sin(\phi)\end{array}\right)=p_{t}^{\mathrm{jet}}\left(\begin{array}{c}
\cos\psi\\
\sin\psi\end{array}\right),\end{equation}
where $\psi$ is the jet hadron direction with respect to the flow
direction $\phi_{\mathrm{flow}}=0$. We have\begin{equation}
\tan\psi=\frac{p_{t}^{\mathrm{hard}}\sin(\phi)}{p_{t}^{\mathrm{soft}}+p_{t}^{\mathrm{hard}}\cos(\phi)}.\end{equation}
Assuming a flat $\phi$ distribution, the probability distribution
for $\psi$ is $(2\pi)^{-1}d\phi/d\psi$, which is in the case of
$p_{t}^{\mathrm{soft}}\ll p_{t}^{\mathrm{hard}}$ given as\begin{equation}
1+\frac{p_{t}^{\mathrm{soft}}}{p_{t}^{\mathrm{hard}}}\cos(\psi),\end{equation}
which should only be considered for $-\pi/2<\psi<\pi/2$, since $\psi$
and $\phi_{\mathrm{flow}}$ have to correspond to the same hemisphere.
We get anisotropies of the order of $p_{t}^{\mathrm{soft}}/p_{t}^{\mathrm{hard}}$,
which means at transverse momenta around $10$~GeV/c, a soft push
as little as 1~GeV/c can produce anisotropies of the order of 10\%.

\section{Coalescence}

For many years, different models have been employed to treat particle
production at different transverse momentum scales. The so-called
intermediate range from 2 to 6 GeV/c has been the domain of coalescence
models \citet{coa1,coa2,coa3,coa4,coa5}, where hadrons are produced
by recombining quarks from the plasma, to be distinguished from {}``fragmentation''
of partons.

In our picture, there are certain aspects which give similar results
as coalescence, but it is not a coalescence approach. Already the
notion {}``intermediate $p_{t}$'' extends to say 20 GeV/c and not
6. The corresponding transverse momentum of hadrons does not originate
from plasma quarks and antiquarks -- the main part is coming from
the original flux tube. Whereas usual flux tube breaking in vacuum
creates quark-antiquark pairs via a tunneling process, the fluid-jet
interaction amounts to replacing these quark-antiquark pairs by partons
from the plasma. So our jet hadrons finally carry {}``some'' transverse
momentum from fluid partons, but only a small fraction. But this is
enough to create for example anisotropies in dihadron correlations.
It will also affect strongly baryon to meson rations, as we are going
to discuss in a separate publication.

\section{Summary}

We presented a theoretical scheme which accounts for bulk matter,
jets, and the interaction between the two. The criterion for bulk-jet
separation is based on parton energy loss. But in addition to the
latter mechanism, there are very important new phenomena which have
not been discussed so far: The interaction between jet hadrons and
soft ones (from fluid freeze-out), and the interaction between the
fluid and jet hadrons at the moment of the creation of the latter
ones. Particle production between zero and (at least) 20 GeV/c is
affected. We understand quantitatively azimuthal anisotropies in single
particle production and dihadron (long range) correlations at large
values of $p_{t}$. 

\begin{acknowledgments}
This research has been carried out within the scope of the ERG (GDRE)
{}``Heavy ions at ultra-relativistic energies'', a European Research
Group comprising IN2P3/CNRS, Ecole des Mines de Nantes, Universite
de Nantes, Warsaw University of Technology, JINR Dubna, ITEP Moscow,
and Bogolyubov Institute for Theoretical Physics NAS of Ukraine. Iu.K.
acknowledges partial support by the State Fund for Fundamental Researches
of Ukraine (Agreement of 2011) and National Academy of Sciences of
Ukraine (Agreement of 2011). K.W. would like to thank Ludmila Malinina
and Konstantin Mikhailov for useful discussions. M.B. thanks the Hessian
LOEWE initiative for financial support.
\end{acknowledgments}
\newpage{}


\begin{thebibliography}{10}
\bibitem{hs06}HS06 is the abbreviation for HEP-SPEC-2006, which is
a benchmark for measuring CPU performances. The machines we use have
an average power of 8.7 HS06 units, this number being referred to
as {}``scaling factor''.
\end{thebibliography}
\end{document}